\documentclass[conference, 10pt]{IEEEtran}
\IEEEoverridecommandlockouts

\usepackage[table]{xcolor}
\usepackage[noadjust]{cite}
\usepackage{fontawesome5}
\usepackage{wasysym}
\usepackage{subcaption}
\usepackage{marvosym}
\usepackage{pdfpages}
\usepackage{multirow}
\usepackage{enumitem}
\usepackage{standalone}
\usepackage{pgfplots}
\usepgfplotslibrary{groupplots, fillbetween}
\usepackage{tikz}
\usetikzlibrary{calc,positioning,fit,shapes}
\usepackage{pbox}

\usepackage[pdftex,bookmarks=false]{hyperref}
\hypersetup{
	pdfsubject = {Hard Shell, Reliable Core: Improving Resilience in Replicated Systems with Selective Hybridization (Extended Version)},
	pdftitle = {Hard Shell, Reliable Core: Improving Resilience in Replicated Systems with Selective Hybridization (Extended Version)},
	pdfauthor = {Laura Lawniczak and Tobias Distler},
	pdfkeywords = {Fault tolerance, State-machine replication, System architectures, Hybridization, Customization, Diversification},
	pdfborder = {0 0 0}
}

\usepackage[nameinlink]{cleveref}

\title{Hard Shell, Reliable Core: Improving Resilience in Replicated Systems with Selective Hybridization}
\IEEEspecialpapernotice{(Extended Version)}

\author{
	\IEEEauthorblockN{Laura Lawniczak and Tobias Distler\thanks{This work was supported by the Deutsche For\-schungs\-ge\-mein\-schaft (DFG, German Research Foundation) – 541017677, 446811880. The paper is an extended version of our SRDS 2025 publication~\cite{lawniczak25hard}. }}
	\IEEEauthorblockA{
		Friedrich-Alexander-Universit\"at Erlangen-N\"urnberg, Germany\vspace*{-2.5mm}
	}
}

\newcommand{\headline}[1]{\vspace{2mm}\noindent\textbf{\textit{#1.}}~}

\usepackage{amsthm}
\newtheorem{propertyrdp}{Property}

\newtheorem{propertyrp}{Property}

\usepackage{xspace}
\newcommand{\shellft}{\textsc{ShellFT}\xspace}

\newcommand{\minas}{\textsc{Minas}\xspace}
\newcommand{\sentry}{\textsc{Sentry}\xspace}
\newcommand{\mentry}{\textsc{Minas+Sentry}\xspace}
\newcommand{\bftsmart}{BFT-SMaRt\xspace}

\usepackage{tabularx}
\newcommand\nocell[1]{\multicolumn{#1}{c|}{}}

\usepackage{listings}
\usepackage{subscript}
\usepackage{xspace}
\usepackage{todonotes}

\font\lsttt=rm-lmtl10 scaled 770
\font\lstbtt=rm-lmtk10 scaled 770
\newcommand{\commentsize}{\fontsize{7.5pt}{0pt}\selectfont}
\newcommand{\pc}{\commentsize\normalfont}

\begin{document}

\bstctlcite{BSTcontrol}

\maketitle

\begin{abstract}
Hybrid fault models are known to be an effective means for enhancing the robustness of consensus-based replicated systems.
However, existing hybridization approaches suffer from limited flexibility with regard to the composition of crash-tolerant and Byzantine fault-tolerant system parts and/or are associated with a significant diversification overhead.
In this paper we address these issues with \shellft, a framework that leverages the concept of micro replication to allow system designers to freely choose the parts of the replication logic that need to be resilient against Byzantine faults.
As a key benefit, such a selective hybridization makes it possible to develop hybrid solutions that are tailored to the specific characteristics and requirements of individual use cases.
To illustrate this flexibility, we present three custom \shellft protocols and analyze the complexity of their implementations.
Our evaluation shows that compared with traditional hybridization approaches, \shellft is able to decrease diversification costs by more than 70\%.
\end{abstract}

\section{Introduction}

\vspace{-.4mm}

Although a variety of causes such as software bugs, hardware errors, or malicious attacks potentially result in arbitrary behavior of components, crash-tolerant state-machine replication protocols~\cite{lamport98part,ongaro14search} today still are the norm in many practical use cases.
Full-fledged Byzantine fault-tolerant~(BFT) protocols~\cite{castro1999practical,distler21byzantine} could offer improvements with regard to resilience, however they are often considered too expensive, both in terms of complexity as well as resource consumption~\mbox{\cite{veronese2011efficient,kapitza12cheapbft,distler16resource}}.
As a consequence, in recent years the use of hybrid fault models has drawn a significant amount of attention due to offering a tradeoff between both worlds~\mbox{\cite{liu16xft,behl17hybrids,decouchant22damysus,messadi22splitbft}}.

In a nutshell, the fundamental idea of hybridization is to combine different fault assumptions within a single system.
For replication protocols, this can for example mean to only tolerate crashes in certain (trusted) components while being resilient against Byzantine faults in other (untrusted) parts~\cite{correia05low,chun07attested}, or to have distinct resilience thresholds for different classes of faults~\mbox{\cite{clement2009upright,porto15visigoth}}.
Unfortunately, despite their effectiveness, existing hybridization approaches have at least one of two drawbacks:
(1)~With the partitioning of trusted/untrusted areas typically being dictated by the protocol design, and resilience thresholds generally applying to the replicated system as a whole, it is inherently difficult to adjust them to new use cases with different demands.
(2)~Requiring each replica to perform a complex set of tasks, they commonly involve a large overhead when it comes to diversifying the replica logic (e.g.,~using N-version programming~\cite{chen78nversion}).

In an effort to address these issues we present \shellft, a novel approach for the hybridization of replicated systems that offers developers an unprecedented degree of flexibility and significantly reduces the costs for diversification.
\shellft builds on the observations that (1)~with regard to overall system robustness, some mechanisms of a replication protocol are more critical or more vulnerable than others, and that (2)~this subset of crucial mechanisms usually depends on the specific use-case scenario of a replicated system.
For example, for user-facing services that are directly accessible from the Internet, the communication with clients~(if not protected properly) can be used as a gateway for attacks.
On the other hand, if a replicated system only serves trusted clients (e.g.,~due to hosting a lower-tier service inside a data center), the leader-replica functionality is often the pivotal part impacting system stability~\cite{clement09making}.
Taking into account these insights, \shellft offers system developers the possibility of deliberately choosing the replication-protocol components and mechanisms that need to be protected against Byzantine faults.
As a \linebreak key benefit, this \emph{selective hybridization} significantly increases overall system resilience at only small additional costs.

To achieves this, \shellft leverages the fine-grained modularization of a replication protocol into clusters of tiny components that each are responsible for a dedicated protocol step.
More specifically, the fact that these clusters are largely independent of each other allows \shellft to treat every cluster as a separate domain that has its own fault model.
Among other things, this for example makes it possible to selectively tolerate a (predefined) maximum number of Byzantine faults in a particular protocol mechanism without the need to tighten overall synchrony assumptions.
Finally, as an additional advantage, the clustering enables a targeted diversification of critical parts which, in contrast to the diversification of entire replicas in traditional hybrid architectures, in a \shellft protocol is inexpensive due to being limited to individual protocol steps.

In particular, this paper makes the following contributions:
(1)~It introduces \shellft, an approach to improve the resilience of state-machine replication protocols by applying selective hybridization.
(2)~It provides details on the \shellft framework, a tool that assists system developers by configuring our \shellft codebase depending on their choices of fault domains.
(3)~It illustrates the flexibility of \shellft by discussing three custom protocols that are tailored to specific use cases.
(4)~It evaluates the three \shellft protocols with regard to complexity, diversification costs, and performance.

\section{Background and Problem Statement}
\label{sec:background}

In this section, we provide necessary background and highlight the differences between the state of the art and \shellft.

\subsection{State-Machine Replication Architectures}
\label{sec:smr-architectures}

\begin{figure}
	\vspace{.9mm}
	\includegraphics{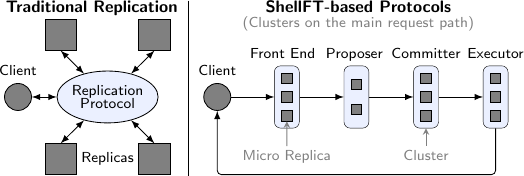}
	\caption{State-machine replication architectures.}
	\label{fig:background}
	\vspace{-1.5mm}
\end{figure}

State-machine replication~\cite{schneider90state} achieves fault tolerance by hosting multiple instances of the same service on different machines~(see Figure~\ref{fig:background}, left side).
To ensure consistency, the replicas run an agreement protocol~\cite{lamport98part,castro1999practical} that establishes a total order on incoming client requests, and in addition performs tasks such as view change and checkpointing.
Traditionally, this multifaceted set of responsibilities is handled by a group of monolithic replicas, however several works have shown that there are alternatives to this basic design.
Specifically, a number of architectures have been proposed that are separating agreement from execution~\mbox{\cite{yin03separating,eischer20spider}}, possibly adding a third stage for the reception of requests~\cite{clement2009upright}.
Although reducing the functionality required by individual nodes compared with the traditional monolithic approach, the scopes of replicas in these architectures still consist of entire protocol stages, including for example the whole consensus process.
Among other things, this leads to significant overhead for the implementation of heterogeneous variants in cases in which N-version programming~\cite{chen78nversion} is applied for diversification.
To some degree, this overhead can be reduced by employing architectures that rely on compartmentalization~\mbox{\cite{whittaker21scaling,messadi22splitbft}}, however for \shellft we strive for an even more fine-grained diversification.

\textit{\underline{Our Approach}:} As basis for the \shellft protocol architecture we rely on micro replication~\cite{distler2023micro}, a concept that splits a replication protocol into atomic tasks (see Figure~\ref{fig:background}, right side) and hence results in replica implementations with very low complexity.
This way, with each micro replica only handling a single protocol step, a tailored diversification of critical protocol parts becomes both feasible and affordable.

Compared with traditional approaches, the distribution of a protocol across a larger number of (small) replicas comes with increased configuration and deployment costs.
For \shellft, we consider this an acceptable trade-off for the fact that micro replication offers us the flexibility to selectively increase the robustness of individual protocol steps.
In general, developing a micro-replicated protocol from scratch is not always straightforward, which is why for \shellft we circumvent this problem by utilizing the already existing Mirador~\cite{distler2023micro} protocol as starting point for our work~(see Section~\ref{sec:base-protocol}).
However, since the general concept behind \shellft's selective hybridization is not intrinsically linked to Mirador, it can be adapted to other micro-replicated protocols once they become available.

\subsection{Hybrid Fault Models}
\label{sec:hybrids}

Although the failures occurring in replicated systems in practice are not only limited to crashes, this does not necessarily mean that resorting to full-fledged Byzantine fault tolerance is automatically the best option, especially when taking into account the additional costs in terms of resources and complexity~\cite{kuznetsov09bftw3,liu16xft}.
Leveraging this observation, several previous works have examined the use of hybrid fault models which (in addition to crashes) assume Byzantine-faulty behavior only in specific system parts and/or under certain circumstances.
In the following, we elaborate on various incarnations of this concept and highlight the differences to our \shellft approach.
Notice that in the context of replicated systems the term ``hybrid'' has been used to describe a plethora of ideas that are orthogonal to the work presented in this paper~(e.g.,~the design of an overall replication protocol as a composition of individual protocols with particular characteristics~\mbox{\cite{cowling06hq, pass17hybrid}}, a management model with dedicated responsibilities~\cite{khan23making}, the combination of failure detection and randomization to solve consensus~\cite{aguilera98failure}).
Thus, below we focus our discussion on approaches that are closely related to our notion of hybridization.

\headline{Special-Purpose Trusted Components}
Applying the concept of architectural hybridization~\cite{correia05low}, the key idea is to split fault assumptions on a spatial level.
Specifically, this means that some parts of a replicated system are considered to be trusted and to only fail by crashing, whereas the remaining parts may be subject to Byzantine faults.
The sizes of these two areas significantly vary between systems.
While some works rely on a comparably large trusted computing base~(e.g.,~a distributed communication subsystem~\cite{correia05low} or a virtualization layer~\cite{reiser07hypervisor,distler11spare}), others have shown that it is possible to reduce the trusted part to relatively small components such as a log~\cite{chun07attested}, a counter~\cite{levin09trinc,veronese2011efficient,kapitza12cheapbft,behl17hybrids}, or other special-purpose modules implementing parts of the replication protocol~\mbox{\cite{decouchant22damysus,messadi22splitbft}}.

As a main benefit, the use of dedicated trusted components enables resource-efficient systems that tolerate up to $f$ Byzantine faults in the untrusted areas by using $2f+1$~replicas, which is the same as for crash-tolerant replication.
On the downside, due to the dependence on special-purpose trusted components there is typically no flexibility with regard to the selection of trusted/untrusted parts.
In all works mentioned above, the partitioning between trusted and untrusted areas is hardwired into the system design, thereby making it inherently difficult to harness these systems for application scenarios in which the fault and threat assumptions differ from the ones they have been developed for.
As an additional drawback, designing replication protocols based on special-purpose trusted components \linebreak is not straightforward, as recent examples have shown~\cite{bessani23vivisecting}.

\textit{\underline{Our Approach}:} In contrast to previous works, \shellft offers the opportunity to freely choose the trusted and untrusted parts of a replicated system at configuration time.
As illustrated in Section~\ref{sec:examples}, this allows systems to be tailored to the individual fault and threat models of specific use cases.

\headline{Relaxed Synchrony Assumptions}
A major reason for the increased complexity and resource consumption of many traditional BFT protocols is rooted in their goal to tolerate Byzantine faults in the presence of an asynchronous network~\cite{castro1999practical}.
Several authors~\cite{porto15visigoth,liu16xft} argue that for many practical use cases such a combination of assumptions is unnecessarily strong; for example, when systems comprise redundant (wide-area) communication links between replicas, and hence make it difficult for an adversary to not only control a subset of replicas but also the network.
XFT~\cite{liu16xft} exploits this insight by designing state-machine replication protocols in such a way that they are able to deal with both a certain number of Byzantine faults as well as asynchrony, but not at the same time.\linebreak
As a main advantage, this approach makes it possible to minimize protocol complexity and keep the minimum number of required replicas to $2f+1$.
On the downside, if indeed the maximum number of Byzantine faults concurs with network partitions, then these kinds of replicated systems do not just lose liveness, but potentially become unsafe.

\textit{\underline{Our Approach}:} \shellft minimizes complexity by adding resilience to selected protocol tasks, not by trading off network asynchrony for Byzantine fault tolerance.
Thus, \shellft protocols remain safe even in the presence of partitions.

\definecolor{darkgreen}{rgb}{0,.5,0}
\def\worstcolor{red!30}
\def\mediumcolor{yellow!30!orange!30}
\def\bestcolor{green!50!darkgreen!30}

\begin{table}
	\vspace{1.8mm}
	\centering
    \setlength{\tabcolsep}{0.74em}
    \renewcommand{\arraystretch}{0.9}
	\begin{tabular}{|l|c|c|c|}
		\cline{2-4}
		\nocell{1} & \multicolumn{2}{c|}{\textbf{Hybridization}} & \textbf{Diversification}\\
		\cline{2-3}
		\nocell{1} & \textbf{Type} & \textbf{Configurability} & \textbf{Granularity}\\
		\hline
		MinBFT~\cite{veronese2011efficient} & Subsystem & \cellcolor{\worstcolor}Hardwired & \cellcolor{\worstcolor}Monolithic replica\\
		XFT~\cite{liu16xft} & Either-or & \cellcolor{\mediumcolor}Global & \cellcolor{\worstcolor}Monolithic replica\\
		UpRight~\cite{clement2009upright} & Fault classes & \cellcolor{\mediumcolor}Global & \cellcolor{\mediumcolor}Protocol stage\\
		VFT~\cite{porto15visigoth} & Fault classes & \cellcolor{\mediumcolor}Global & \cellcolor{\worstcolor}Monolithic replica\\
		\hline
		\textbf{\shellft} & Fault domains & \cellcolor{\bestcolor}Selective & \cellcolor{\bestcolor}Protocol step\,/\,task\\
		\hline
	\end{tabular}
	\caption{Comparison of hybridization approaches.}
	\label{tab:background}
	\vspace{-2mm}
\end{table}

\headline{Distinction of Fault Classes}
Systems such as UpRight~\cite{clement2009upright} distinguish between two thresholds to determine the overall number of replicas $n = 2u + r + 1$: a threshold~$u$, which denotes the maximum number of tolerated faults in total (i.e.,~crashes plus Byzantine faults), and a threshold~$r$ representing the maximum number of tolerated Byzantine faults.
VFT~\cite{porto15visigoth} extends this idea of handling some fault classes separately by introducing further thresholds for slow replicas and correlated faulty behavior.
Similarly, other works use thresholds for disconnected sites~\cite{khan23making} or concurrently recovering replicas~\cite{sousa10highly}.

Distinguishing between fault classes offers the advantage of reducing replication costs when not all expected faults are assumed to be of arbitrary nature~\cite{thambidurai88interactive}.
On the other hand, as applied by UpRight and other systems, the method is rather coarse-grained due to defining fault assumptions at the level of the entire replicated system.
That is, even though UpRight comprises three stages,
the threshold values are cross-cutting parameters and thus rule out stage-specific configurations.

\textit{\underline{Our Approach}:} \shellft separates a replication protocol into loosely coupled clusters (each representing an individual protocol step) and allows each cluster to be flexibly assigned its own fault model.
Within such a fault domain, applying \linebreak different thresholds for different fault classes would be possible, however the specifics are outside the scope of this paper.

\subsection{Problem Statement}

As summarized in Table~\ref{tab:background}, our analysis in the previous sections has shown state-of-the-art approaches to have two main drawbacks:
(1)~A limited flexibility with respect to hybridization, either because the partitioning between trusted and untrusted parts is hardwired into the system design, or due to fault classes being configured globally.
(2)~A high diversification overhead caused by the concentration of complex functionality at replicas that are required to handle a whole protocol stage or even the entire replication protocol.
In the following section, we elaborate on how \shellft addresses these \linebreak issues by offering selective hybridization and diversification.

With regard to diversification, we especially focus on measures that introduce heterogeneity by implementing replicas in different programming languages~\cite{chen78nversion}, and thereby reduce the risk of vulnerabilities in the language runtime or libraries affecting multiple replicas and therefore, in the worst case, the whole replicated system.
Due to the programming effort for these kinds of approaches (and hence the associated economical cost) typically depending on the complexity of the replica logic, our goal is to improve this situation by minimizing the amount of code that actually needs to be diversified in order to increase the robustness of selected protocol steps.
\section{\shellft}
\label{sec:shellft}

\shellft is both a novel concept for flexibly applying hybridization in replicated systems as well as an accompanying framework that automates the tailoring to specific use cases.
In this section, we first give an overview of the main idea behind \shellft and then provide details on the customization.

\subsection{Overview}
\label{sec:shellft-overview}

\shellft relies on a system architecture in which each replication-protocol task is handled by a dedicated cluster of micro replicas.
As a key benefit, this partitioning allows us to treat each of these clusters as a separate fault domain with individual fault and threat model.
Specifically, we distinguish between three domain types: \emph{Shell} clusters are considered to require resilience against arbitrary faults, \emph{core} clusters represent the crash-tolerant parts of the system, and \emph{filter} clusters act as a barrier between these two, thereby shielding the core from the shell.
More precisely, the three domain types have the following characteristics:

\begin{itemize}[leftmargin=5mm]
	\item \textbf{Shell:} Replicas of the clusters belonging to this type of domain may be subject to Byzantine faults.
	The assignment of clusters to this category is made by the user of the \shellft framework based on the individual properties \linebreak and requirements of a protocol's application scenario.
	\item \textbf{Filter:} Replicas of this domain type are assumed to only fail by crashing, however they receive inputs from at least one shell cluster and therefore require means to tolerate Byzantine-faulty input values provided by these sources.
	The classification as filter cluster is automatically made by the \shellft framework based on knowledge of the identity of the user-defined shell clusters as well as the cluster-interaction dependencies of a replication protocol.
	\item \textbf{Core:} Replicas included in the core domain are only subject to crash faults and exclusively process inputs obtained from other potentially crash-faulty clusters (i.e.,~filters and other cores).
	The \shellft framework automatically labels all clusters as cores that are neither shells nor filters.
\end{itemize}

\begin{figure}[b!]
	\vspace{-.2mm}
    \centering
    \includegraphics{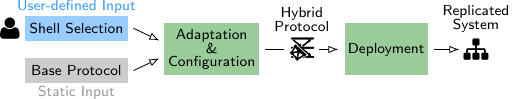}
    \caption{Overview of the \shellft tailoring process.}
    \label{fig:approach}
\end{figure}

\noindent{}As illustrated in Figure~\ref{fig:approach}, as a starting point for the tailoring process \shellft relies on the implementation of a purely crash-tolerant \emph{base protocol} which our framework then successively transforms into a hybrid protocol implementation taking into account a user's shell selection.
Among other things, this process may include changes to the number of replicas comprised in certain clusters, a reconfiguration of the thresholds based on which individual replicas accept an input value, or the replacement of crash-tolerant protocol mechanisms with more resilient Byzantine fault-tolerant logic.
Once the tailoring is complete, the result is a custom and preconfigured \linebreak hybrid protocol implementation that is ready for deployment.

To account for the specific characteristics of our three fault domains, when deploying a \shellft system we physically isolate the shell from the rest of the protocol by running the shell clusters on a separate group of machines.
This way, if a Byzantine fault (e.g.,~caused by an attack) is not limited to a shell replica but compromises the entire machine hosting the replica, the filter and core clusters still remain unaffected.
In sum, the use of two server groups enables us to minimize \shellft's resource consumption and deployment costs, \linebreak while at the same time offering a high degree of resilience.

\subsection{Base Protocol}
\label{sec:base-protocol}

The base protocol is a crash-tolerant replication protocol that we specially developed as basis for our tailoring process.
To facilitate the tailoring, we designed the base protocol in such a way that its architecture and workflow closely resemble the architecture and workflow of the existing micro-replicated BFT protocol Mirador~\cite{distler2023micro}.
Most notably, this allows us to substitute selected base-protocol clusters for their Mirador counterparts as part of the adaptation~(see Section~\ref{sec:shellft:framework}).

For the agreement on client requests, the base protocol relies on a consensus algorithm that is comparable to Paxos~\cite{lamport98part}.
In particular, we target systems in which the number of concurrent failures to tolerate is small (e.g., $f \leq 2$).
Such system environments match our goal of further strengthening robustness through diversification, which in practice (despite the benefits offered by \shellft) is unlikely to become affordable for deployments comprising tens or even hundreds of servers. 
For the base protocol, this means that scalability with the number of faults~($f \gg 1$) is not a requirement.
Exploring our approach in large systems is a potential direction for future work and \linebreak presumably involves the development of another base protocol.

Although the base protocol is executable on its own, we did not concentrate on optimizing for such a scenario.
Instead, our primary focus was to create a parameterized template protocol that the \shellft framework can use to support selective hybridization.
With the resulting tailored protocol later potentially being subject to Byzantine faults, this for example means that replicas in the base protocol already communicate via authenticated messages to prevent adversaries from successfully impersonating correct replicas.

The full specification and system model of the base protocol can be found in Appendix~\ref{sec:specification}.
In the following, we give a brief overview of the protocol and its parameterization options.

\begin{figure}
    \centering
    \includegraphics{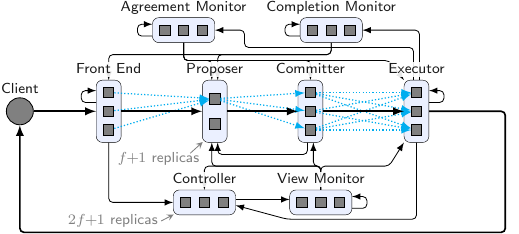}
    \caption{Base protocol.}
    \label{fig:shellft-base}
	\vspace{-3.5mm}
\end{figure}

\vspace{-.9mm}

\headline{Protocol Architecture}
As shown in Figure~\ref{fig:shellft-base}, the base protocol consists of eight micro-replica clusters, four of which represent the main request path through the system.
Specifically, incoming client requests first arrive at a cluster of front-end replicas.
Next, they are forwarded through a chain of proposer, committer, and executor clusters which together implement a Paxos-style consensus process~\cite{kirsch08paxos} that assigns a unique sequence number to each request.
Finally, the executors process committed requests in the order of their sequence numbers and send the corresponding replies back to the client.

In addition to these four main clusters, the base protocol contains three different monitor clusters that are responsible for obtaining and distributing progress information on the consensus process (agreement monitor), executed requests (completion monitors), and view number (view monitors).
Acting as control loops, monitors enable replicas of the main path to perform garbage collection of no longer needed state (e.g., agreement slots that are superseded by a stable checkpoint).
For clarity, we omit a fine-grained discussion of these mechanisms because (1)~control loops in the base protocol follow the same principles as control loops in Mirador and (2)~their specifics (like other Mirador details deliberately left out here) are not of relevance for the contributions of this paper, namely selective hybridization and diversification.
The same applies to the controller cluster, a group of replicas that monitors agreement and execution progress and if necessary triggers a view change for proposers, committers, and executors.

Relying on this architecture, information flows through the system by replicas of a cluster providing the outcomes of their own protocol step as inputs to the replicas of their successor cluster(s), typically in an all-to-all manner, as illustrated for the main-path clusters in Figure~\ref{fig:shellft-base} (dotted blue arrows).
Deciding on a protocol step in general requires a replica to analyze the input values provided by different predecessors and accept a value once a predefined threshold is reached; either in terms of quorum size (e.g.,~an executor commits a request once \mbox{$f+1$}~committers have confirmed the request's reception) or in terms of numerical value (e.g.,~the current view is determined as the \mbox{$f+1$} highest number announced by view monitors).

\headline{Parameterization}
Leveraging its micro-replicated protocol architecture, the base protocol serves as starting point for the creation of tailored replication protocols.
In particular, the base protocol offers adaptations in three main dimensions:

\begin{itemize}[leftmargin=5mm]
	\item \textbf{Adjustment of Replication Factors:} With clusters being largely independent of each other, the number of involved replicas can be defined on a per-cluster basis.
	For some protocol steps, this offers the opportunity to achieve robustness against Byzantine faults by adding replicas to the corresponding cluster~(see Section~\ref{sec:examples}).
	\item \textbf{Configuration of Acceptance Thresholds:} Raising the bar for the acceptance of input values enables a replica to limit the impact faulty inputs can have on its decisions.
	Usually, such a measure is complemented by a matching increase in the replication factor of the respective predecessor cluster.
	\item \textbf{Substitution of Protocol Mechanisms:} Exploiting the loose coupling of micro-replica clusters, entire mechanisms of the base protocol can be modularly replaced by their counterparts from other micro-replicated protocols.
	As shown in Section~\ref{sec:sentry}, this for example makes it possible to substitute the base protocol's logic for distributing proposals with an enhanced mechanism that is resilient against equivocation by a faulty proposer.
\end{itemize}

\noindent{}Unlike the selection of shell clusters, decisions regarding base-pro\-to\-col adaptions are not made by users, but by the \shellft framework as part of the automated tailoring process.

\subsection{Tailoring Process}
\label{sec:shellft:framework}

The main purpose of the tailoring process is to transform the base protocol into a customized implementation.
To automate this process and set up the resulting system, we developed the \shellft framework.
Next, we present the tailoring in detail.

\headline{Cluster Adaptation}
Starting with the user-specified selection of shell clusters, the \shellft framework first decides on which functionality and clusters are required for the chosen shell configuration.
For this purpose, the framework relies on a predefined set of rules that for each base-protocol cluster defines the specific actions that need to be performed in order for the cluster to become a shell.
The extent of these actions depends on the particular protocol step affected and hence varies between clusters~(cf.~Section~\ref{sec:shellf:transformation}).
As summarized in Table~\ref{tab:adaptation}, for most clusters the base-protocol implementation can either be directly used without modification, or substituted in place with its counterparts from the Byzantine fault-tolerant protocol Mirador.
In other cases, the changes are more wide-ranging and, for example, affect multiple clusters.
Most notably, putting the proposer inside a shell domain leads to the addition of new clusters, as further detailed in Section~\ref{sec:sentry}.

\begin{table}
	\vspace{1.8mm}
	\centering
    \setlength{\tabcolsep}{0.48em}
    \renewcommand{\arraystretch}{0.85}
	\begin{tabular}{|l|c|c|}
		\hline
		\multicolumn{1}{|c|}{\textbf{Shell Selection}} & \multicolumn{1}{c|}{\textbf{Replacements}} & \multicolumn{1}{c|}{\textbf{Size Update}}\\
		\hline
		Front end & \multicolumn{1}{c|}{--} & \multicolumn{1}{c|}{--}\\
		Proposer & Mirador agreement stage & \multicolumn{1}{c|}{--}\\
		Committer & Adapted proposer, Mirador executor & $3f+1$\\
		Executor & Mirador executor and client & $3f+1$\\
		\hline
		Agreement monitor & Mirador agreement monitor & $3f+1$\\
		Completion monitor & Mirador completion monitor & $3f+1$\\
		View monitor & Mirador view monitor & $3f+1$\\
		\hline
		Controller & \multicolumn{1}{c|}{--} & \multicolumn{1}{c|}{--}\\
		\hline
	\end{tabular}
	\vspace{-1mm}
	\caption{Cluster replacements and size updates performed by the \shellft framework if a cluster is selected as shell.}
	\label{tab:adaptation}
	\vspace{-2.5mm}
\end{table}

\headline{Cluster Configuration}
At this point, the \shellft framework knows all relevant clusters and which of them are part of the shell domain.
Using this information, in the next step the framework classifies the remaining clusters as either filters or cores based on a dependency graph that models the interaction between clusters (cf.\ Figure~\ref{fig:shellft-base}).
Most importantly, as soon as a cluster processes direct input from shell functionality, it is automatically included in the filter domain.
This rule also applies if multiple inputs are combined and only some of them come from the shell.
Only clusters that receive inputs \textit{solely} from filters or cores (i.e.,~replicas that the framework user assumes to \linebreak fail by crashing) are themselves assigned to the core domain.

Once all clusters are assigned a domain, the \shellft framework then proceeds by configuring the individual size of each cluster.
With filter and core clusters representing crash-tolerant domains, their replication factors remain the same as in the base protocol.
In contrast, to account for potential Byzantine faults, shell clusters are typically expanded by $f$~additional replicas, resulting in BFT-typical cluster sizes of $3f+1$.
To account for changes in replication factors, for those shell clusters whose sizes have increased (in case there are any), the \shellft framework in a final configuration step increases the corresponding acceptance thresholds of all affected (shell and filter) successor clusters by~$f$.
This allows replicas of the successor clusters to make effective use of the $f$~additional inputs that are now available as a result of the cluster expansion, and to thereby actually tolerate Byzantine faults in the shell.
For more details on how specific transformation decisions are made, please refer to Section~\ref{sec:shellf:transformation}.

\begin{figure}
    \includegraphics{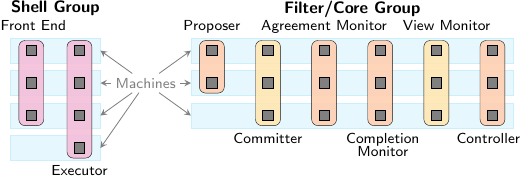}
	\vspace{-5.5mm}
    \caption{Group-based deployment on separate sets of machines.}
    \label{fig:deployment}
	\vspace{-2mm}
\end{figure}

\headline{Cluster Deployment}
\label{sec:shellft:deployment}
Although clusters are isolated entities from a protocol perspective, there are several opportunities with regard to their deployment; they range from hosting each replica on a separate server to co-locating replicas of different clusters within the same thread.
For \shellft we exploit this flexibility to achieve a balance between resilience and efficiency by (1)~physically separating shell clusters from the rest of the system and (2)~integrating multiple replicas with each other whenever possible.
This strategy results in a setting comprising two groups of machines: one group for shell replicas and one group for filter and core replicas.
Figure~\ref{fig:deployment} shows an example of such a deployment for a configuration in which the shell consists of front ends and executors~(cf.\ Section~\ref{sec:minas}).
Keeping the shell clusters isolated from the crash-tolerant domains increases robustness because, even if an adversary manages to compromise an entire machine of the shell group, the adversary still does not have control over filter and core replicas.
At the same time, without impairing availability, the co-location of replicas from different clusters within each of the two groups minimizes the communication overhead between the corresponding protocol steps and therefore improves both resource consumption and efficiency.

\subsection{Pattern-based Protocol Transformation}
\label{sec:shellf:transformation}

During its adaptation and configuration steps, the \shellft tailoring process leverages the fact that micro-replicated protocols are designed as a composition of established architectural patterns with specific safety and liveness guarantees~\cite{distler2023micro}.
This way, the transformation essentially becomes the task of translating a pattern that provides a certain property in the presence of crash faults into its counterpart pattern providing the same property in the presence of Byzantine faults.

\begin{figure}[b!]
	\vspace{-.5mm}
	\begin{center}
		\subfloat[Reliable distribution pattern\label{fig:transformation-reliable}]{
			\includegraphics[page=1]{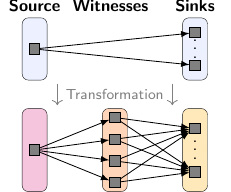}
		}
		\hspace{2mm}
		\subfloat[Relay pattern\label{fig:transformation-relay}]{
			\includegraphics[page=2]{figures/patterns.pdf}
		}
	\end{center}
	\vspace{-3mm}
	\caption{Transformation based on design patterns.}
    \label{fig:transformation}
\end{figure}

In this context, it is important to note that \shellft's tailoring process does not change how individual patterns are interweaved to form the overall protocol composition, and thus ensures that the associated inter-pattern correctness arguments remain unaffected.
Instead, the transformation performed by \shellft occurs at the pattern level, thereby making it significantly easier to maintain correctness due to only a comparably small number of precisely defined properties having to be preserved across the transformation.
Next, we illustrate this ap-\linebreak{}proach for the base protocol's two main patterns (see Figure~\ref{fig:transformation}).

\headline{Reliable Distribution Pattern}
This pattern propagates a value from a potentially faulty \emph{source} replica (possibly via intermediate replicas called \emph{witnesses}) to a group of \emph{sink} replicas while ensuring that correct sinks do not accept diverging values.
Similar to traditional protocols~\mbox{\cite{kirsch08paxos,castro1999practical}}, such a task, for example, marks the first step of the base protocol's agreement logic.
For correctness, two properties are required from this pattern and, as proven in Appendix~\ref{sec:proofs-rdp}, both the base version \linebreak and the transformed version of the pattern provide them:

\vspace{-1.3mm}

\begin{propertyrdp}
	\label{the:reliable-distribution-1}
	If a correct sink~$s_1$ accepts a value~$v$ and another correct sink~$s_2$ accepts a value~$v'$, then $v = v'$.
\end{propertyrdp}

\vspace{-3.5mm}

\begin{propertyrdp}
	\label{the:reliable-distribution-2}
	If the source is correct and proposes~$v$, all correct sinks eventually accept~$v$, even if $f$ witnesses are faulty.
\end{propertyrdp}

\vspace{-1.3mm}

As shown at the top of Figure~\ref{fig:transformation-reliable}, the (crash-fault) base version of this pattern involves the source directly broadcasting the proposed value to all sinks.
In contrast, if the source cluster is selected as shell, \shellft's tailoring process switches to the Byzantine-resilient counterpart pattern~(see bottom of Figure~\ref{fig:transformation-reliable}).
Here, an additional cluster of $3f+1$~witness replicas observe the value proposed by the source, and sinks only accept a value after having received matching opinions from $2f+1$~witnesses.
That is, in this case the transformation includes (1)~the insertion of a new cluster as well as (2)~the increase of the sinks' acceptance threshold from $1$ to $2f+1$.

\headline{Relay Pattern}
This pattern is used to distribute a value~(e.g., the sequence number of a reached checkpoint) from multiple sources to multiple sinks; up to $f$~sources may be faulty.
A correct source either puts out a specific value~$v$ or no value at all.
Under these conditions, the relay pattern ensures that if a correct sink accepts~$v$, then eventually all correct sinks accept \linebreak the same value.
In the base protocol, such a guarantee represents the foundation of control loops~(cf.\ Section~\ref{sec:base-protocol}).
As proven in Appendix~\ref{sec:proofs-rp}, \shellft's tailoring process maintains two key properties across the protocol transformation:

\vspace{-1.3mm}

\begin{propertyrp}
	\label{the:relay-1}
	If a correct sink accepts a value~$v$, then $v$ was proposed by a correct source.
\end{propertyrp}

\vspace{-3.5mm}

\begin{propertyrp}
	\label{the:relay-2}
	If a correct sink accepts a value~$v$, then all correct sinks eventually accept~$v$, even if $f$ relays are faulty.
\end{propertyrp}

\vspace{-1.8mm}

In the base pattern~(Figure~\ref{fig:transformation-relay}), \mbox{$2f+1$} sources send their value to a cluster of $2f+1$ \emph{relays}.
In addition, relays propagate an accepted value among each other.
A correct relay accepts a value based on either $t_s = f+1$~matching inputs from different sources or one accepted value forwarded by another relay.
A sink accepts a value after obtaining $f+1$~matching inputs from different relays.
If the sources are in the shell, our transformation (1)~changes the the source-cluster size from $2f+1$ to \mbox{$3f+1$}~replicas and (2)~increases $t_s$ to $2f+1$~inputs.

\subsection{Selective Diversification}
\label{sec:shellf:diversification}

For additional resilience, \shellft enables selective hybridization to be complemented with \emph{selective diversification}.
In general, the diversification of replicas (i.e.,~the use of heterogeneous implementations or deployments) significantly lowers the probability of common-mode failures and makes it more difficult for an adversary to take over multiple replicas at once~\cite{gashi2004designing,garcia14analysis,garcia19lazarus}.
\shellft improves this process in two complementary ways: It reduces the diversification costs for critical \linebreak parts and, at the same time, increases the diversification space.

\vspace{-1mm}

\headline{Reducing the Diversification Costs}
While monolithic architectures suffer from high diversification costs, \shellft allows diversification techniques to be applied to individual clusters, and hence with low overhead.
Having been identified as the most critical/vulnerable parts of the overall system, the shell clusters are primary candidates for methods such as N-version programming.
However, if considered beneficial, additional filter and core clusters may be included to further improve resilience.
Besides, our approach complements currently \linebreak developed LLM-based automation techniques for N-version programming~\cite{ron2024gal}, which require isolated (``pure'') functions.

\begin{table}
	\vspace{1.8mm}
	\centering
    \setlength{\tabcolsep}{0.43em}
    \renewcommand{\arraystretch}{0.85}
	\begin{tabular}{|l|r|r|r|r|r|r|}
		\cline{2-7}
		\nocell{1} & \multicolumn{6}{c|}{\textbf{\# Successful Exploits}} \\
		\cline{2-7}
		\nocell{1} & \multicolumn{1}{c|}{1} & \multicolumn{1}{c|}{2} & \multicolumn{1}{c|}{3} & \multicolumn{1}{c|}{4} & \multicolumn{1}{c|}{5} & \multicolumn{1}{c|}{9} \\
		\hline
		Monolithic protocol & \cellcolor{\bestcolor}0\% & \multicolumn{5}{c|}{\cellcolor{\worstcolor}100\%} \\
		Group-based deployment & \cellcolor{\bestcolor}0\% & \cellcolor{yellow!28.4!orange!30}42.9\% & \multicolumn{4}{c|}{\cellcolor{\worstcolor}100\%} \\
		Fully diversified base protocol & \cellcolor{\bestcolor}0\% & \cellcolor{yellow!50!white!30}12.5\% & \cellcolor{yellow!30}25\% & \cellcolor{yellow!50!orange!30}37.5\% & \cellcolor{orange!30}50\% & \cellcolor{\worstcolor}100\% \\
		\hline
	\end{tabular}
	\vspace{-1mm}
    \caption{Probability of system-wide failure ($f=1$).}
    \label{tab:diversification_crash_perc}
	\vspace{-3.5mm}
\end{table}

\vspace{-.6mm}

\headline{Increasing the Diversification Space}
As diversification space we define the number of possible heterogeneous configurations deployed within a system.
A larger diversification space allows to use a more diverse system layout and hence increases the system's resilience as a whole by minimizing the number of individual parts that are affected by a single common-mode failure (e.g., an exploit or bug in the runtime environment).
In traditional monolithic implementations, each replica runs on exactly one physical node using one configuration. This results in the diversification space being directly related to the replication factor dependent on $f$.
In contrast, in \shellft a replica is a logical construct that is more independent of the physical \linebreak layout, which offers the opportunity to increase the diversification space without having to adjust the replication factor.

To illustrate this aspect, \Cref{tab:diversification_crash_perc} shows how many successful exploits an attacker requires to disrupt a diversified system. Using a heterogeneous monolithic protocol, each exploit takes down one replica, and with the second exploit (for $f=1$), the system is no longer operable.
In contrast, \shellft's group-based deployment presented in \Cref{fig:deployment} already increases overall system resilience. With a \shellft-based system, an attacker first has to find exploits for $f+1$ micro replicas of the same type. If the attacker has no intricate knowledge of the system deployment that would allow the attacker to target parts of the system specifically, this decreases the percentage of a system-wide failure to less than 50\% for two exploits.

Note that if resilience is the main goal, this can be extended to a point where all micro replicas are run in their own deployment configuration and diversified individually. Such \textit{full diversification} significantly decreases the likelihood of a system-wide failure, with only 50\% for even five exploits.
\section{\shellft Protocol Examples}
\label{sec:examples}

\shellft enables replicated systems that are tailored to the individual characteristics and requirements of use cases.
To illustrate how different shell selections influence the resulting protocol, in this section we present three \shellft protocols produced by our framework.
\minas is based on the concept of perimeter security and splits the protocol into outside and inside clusters.
In \sentry, the shell consists of the protocol steps that are most critical for the overall system.
Our third protocol is a composition of \minas and \sentry showing that different shell selections can be combined with each other.

\subsection{\minas: Perimeter Security}
\label{sec:minas}

\minas applies the principle of perimeter security~\cite{avolio1994network,moubayed2019software}. Of our three protocols, it best represents the visual image of a separation between an outer shell and an inner core.

\vspace{-.7mm}

\headline{Use-Case Scenario}
\minas targets data-center environments in which all replicas are connected via a private network that is isolated from public traffic. Such a scenario is not only common for local replicated services but can also be found in hierarchical geo-replicated systems~\cite{amir10steward,calder11mas,eischer20spider}.
Combined with perimeter security inside a data center (e.g., multiple firewalls, network segmentation~\cite{mhaskar2021formal}, SDNs~\cite{benzekki2016software}, intrusion detection~\cite{mukherjee1994network,falcao2019quantitative}), this means that only client-facing clusters must reside in the outer perimeter; all others can be placed in an \linebreak inner perimeter and are thus less exposed to malicious attacks.

\vspace{-.7mm}

\headline{Selection of Shell Functionality}
Given these properties, we instruct the \shellft framework to put the two client-facing clusters~(i.e.,~front ends and executors) inside the shell.

\vspace{-.7mm}

\begin{figure}[b!]
	\vspace{-1mm}
    \centering
    \includegraphics{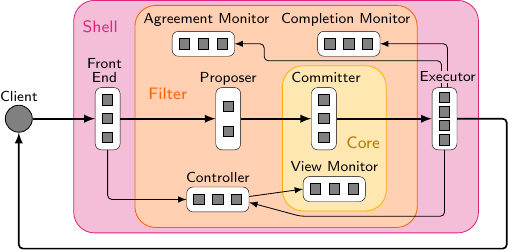}%
	\vspace{-1mm}
    \caption{Overview of the \minas system architecture; some cluster dependencies have been omitted for better readability.}
    \label{fig:minas}
\end{figure}

\headline{Protocol Architecture}
Figure~\ref{fig:minas} shows the resulting \minas system architecture after the tailoring process.
As indicated in Table~\ref{tab:adaptation}, assigning the executor cluster to the shell results in replacements of the executor and client logic, whereas the selection of the front-end cluster does not require any changes.
Overall, with only two clusters in the shell, most parts of the system remain in the filter and core domains, thereby keeping replication costs close to those of the original crash-tolerant base protocol while significantly increasing resilience.

\vspace{-.7mm}

\headline{Additional Details}
As shown by previous works~\cite{yin03separating,clement2009upright}, it is possible to design BFT execution stages requiring only $2f+1$~replicas.
Applying this optimization to the executor cluster is outside the scope of this paper, but would further minimize \minas's resource consumption.
In particular, with regard to the deployment of the system~(see Section~\ref{sec:shellft:framework}, ``Cluster Deployment''), it would allow \minas to reduce the size of its shell group from $3f+1$ to $2f+1$~machines.

\subsection{\sentry: Safety First}
\label{sec:sentry}

For the second protocol \sentry, we take a different ap\-proach. In \sentry, security considerations do not include the locations of individual system parts, instead shell replicas are selected based on the negative impact that a cluster can have on the overall system when exhibiting Byzantine behavior.

\headline{Use-Case Scenario}
Several recent works argue that, given the choice, for many replicated systems in practice it is much more important to preserve safety than liveness~\cite{porto15visigoth,messadi22splitbft}.
The rationale behind this consideration is the insight that from a client perspective, for example, it is usually better to receive no response at all than to receive an erroneous response that reflects inconsistent system state.
In addition, it is generally easier to automatically detect liveness issues (e.g.,~by using external monitoring tools) than safety violations.
Following this idea, for \sentry we specifically apply selective hybridization to those parts of a replicated system that are critical for safety.

\headline{Selection of Shell Functionality}
In a crash-tolerant protocol, there are typically four tasks that pose a particular threat to the safety if they are subject to Byzantine faults:
(1)~Faced with an agreement protocol that consists of only two phases, by performing equivocation (i.e.,~proposing different client requests for the same sequence number to different follower replicas) a Byzantine leader replica can trick correct followers into executing requests in diverging orders.
This in turn may cause replica states to become inconsistent.
(2)~In a similar way, a Byzantine leader is able to manipulate the outcome of a view change by distributing different opinions on the set of requests that need to be re-proposed in the new view.
(3)~Having executed a request, a Byzantine replica may provide the client with an incorrect result.
(4)~If a state transfer becomes necessary (e.g.,~due to a replica rejoining the rest of the system after the end of a temporary network partition), a Byzantine replica can supply a correct replica with an incorrect checkpoint and thereby tamper with the correct replica's state.

In our base protocol, these four tasks are the responsibilities of two clusters, proposer and executor, which is why for \sentry we select both of them as shell.
By design, faulty replicas of all other clusters can only impede the liveness of the overall protocol, even if they show Byzantine behavior.

\begin{figure}
    \centering
    \includegraphics{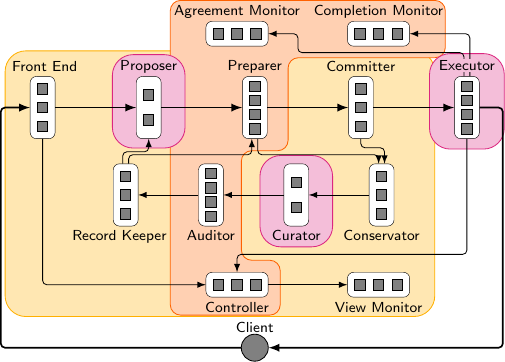}%
	\vspace{-2mm}
    \caption{Overview of the \sentry system architecture; some cluster dependencies have been omitted for better readability.}
    \label{fig:sentry}
\end{figure}

\headline{Protocol Architecture}
Based on this configuration, the \shellft framework produces the system architecture shown in Figure~\ref{fig:sentry}.
As the most significant change, in consequence of the proposer being part of the shell, \shellft's tailoring process applies the replacements associated with the re\-li\-able-dis\-tri\-bu\-tion pattern~(see Section~\ref{sec:shellf:transformation}) by substituting the proposer and committer clusters with the agreement stage of Mirador, thereby introducing an additional preparer cluster on the \linebreak main request path. This enables \sentry to deal with equivocation attempts potentially made by a Byzantine proposer.

Furthermore, off the main path, the agreement stage now includes Mirador's clusters for performing a view change, which in a nutshell works as follows.
Once a view change is triggered, a specific cluster (``conservators'') collects the preparers' and committers' opinions on the requests to be re-proposed in the new view.
Led by a dedicated replica (``curator''), this information is then agreed on using a three-step consensus process, which is similar to the three-phase agreement on the main path.
To complete the view change, at the end the \linebreak agreed outcome is fed back to the proposer and preparer clusters, enabling them to continue with normal-case operation.

To summarize, in \sentry's architecture the base protocol's proposer functionality is divided among two clusters, with the \sentry proposer handling the main request agreement and the curator leading the view change.
As shown in Figure~\ref{fig:sentry}, this results in the curator also constituting a part of the shell.
For the third shell cluster, the executor, the \shellft's tailoring entails the same modifications and benefits as in \minas.

\subsection{\mentry}

For our third protocol, we combine \minas and \sentry to \mentry in order to illustrate that there is no need to focus on a single criterion when selecting the shell functionality.
In recent years, the concepts of zero trust~\cite{mehraj2020establishing} or the rogue administrator~\cite{claycomb2012insider} have received much attention in the area of cloud computing.
So even if a deployment environment fits the characteristics of \minas, for certain applications additional safety mechanisms can still be beneficial.
The resulting system architecture of \mentry is very similar to \sentry{}'s architecture in \autoref{fig:sentry}, except that the front-end cluster in \mentry is also part of the shell.
With this design, \mentry is able to target two threat vectors: The shell now contains both the most \textit{exposed} functionality from \minas as well as the most \textit{critical} functionality as defined in \sentry.

\section{Diversification-Cost Analysis}

Of the two main goals of our work, selective hybridization and selective diversification, the former is inherently provided by \shellft's design~(as discussed in Section~\ref{sec:shellft}).
For this reason, in this section we focus on assessing the diversification costs of \shellft-based replicated systems.
\begin{table*}
	\centering
    \setlength{\tabcolsep}{0.42em}
    \renewcommand{\arraystretch}{0.85}
    \begin{tabular}{|l|l||c|c|c|c|c|c|}
        \cline{3-8}
        \nocell{2} & \multicolumn{6}{c|}{\textbf{Protocol}} \\\hline
        \textbf{Functionality Component} & \textbf{Cluster Counterpart} & \textbf{Baseline} & \textbf{Hybrid} & \textbf{Mirador} & \textbf{\minas} & \textbf{\sentry} & \textbf{\mentry}\\\hline

        Accept requests from clients & Front end & $2f+1$ & \cellcolor{black!75}\color{white}$2f+1$ & \cellcolor{black!75}\color{white}$2f+1$  & \cellcolor{black!75}\color{white}$2f+1$ & $2f+1$ & \cellcolor{black!75}\color{white}$2f+1$ \\
        Assign sequence numbers to requests & Proposer & $2f+1$ & \cellcolor{black!75}\color{white}$2f+1$ & \cellcolor{black!75}\color{white}$\hphantom{2}f+1$   & $\hphantom{2}f+1$           & \cellcolor{black!75}\color{white}$\hphantom{2}f+1$  & \cellcolor{black!75}\color{white}$\hphantom{2}f+1$  \\
        Verify and relay proposals & Preparer & $0$ & $0$ & \cellcolor{black!75}\color{white}$3f+1$  & $0$             & $3f+1$          & $3f+1$          \\
        Confirm sequence-number assignment & Committer & $2f+1$ & \cellcolor{black!75}\color{white}$2f+1$ & \cellcolor{black!75}\color{white}$3f+1$ & $2f+1$          & $2f+1$          & $2f+1$          \\
        Execute requests (and send/store reply) & Executor & $2f+1$ & \cellcolor{black!75}\color{white}$2f+1$ & \cellcolor{black!75}\color{white}$3f+1$  & \cellcolor{black!75}\color{white}$3f+1$ & \cellcolor{black!75}\color{white}$3f+1$ & \cellcolor{black!75}\color{white}$3f+1$ \\\hline

        Monitor progress (to trigger view change) & Controller & $2f+1$ & \cellcolor{black!75}\color{white}$2f+1$ & \cellcolor{black!75}\color{white}$2f+1$    & $2f+1$      & $2f+1$          & $2f+1$          \\
        Monitor and broadcast current view & View monitor & $2f+1$ & \cellcolor{black!75}\color{white}$2f+1$ & \cellcolor{black!75}\color{white}$3f+1$  & $2f+1$      & $2f+1$          & $2f+1$          \\
        Gather prepared requests for next view & Conservator & $0$ & $0$ & \cellcolor{black!75}\color{white}$3f+1$   & $0$         & $2f+1$          & $2f+1$          \\
        Decide on unique set of requests (across views) & Curator & $0$ & $0$ & \cellcolor{black!75}\color{white}$\hphantom{2}f+1$        & $0$         & \cellcolor{black!75}\color{white}$\hphantom{2}f+1$  & \cellcolor{black!75}\color{white}$\hphantom{2}f+1$  \\
        Verify set of requests to re-propose & Auditor & $0$ & $0$ & \cellcolor{black!75}\color{white}$3f+1$       & $0$         & $3f+1$          & $3f+1$          \\
        Store and provide set to re-propose & Record keeper & $0$ & $0$ & \cellcolor{black!75}\color{white}$3f+1$ & $0$         & $2f+1$          & $2f+1$          \\\hline

        Determine active consensus instances & Agreement monitor & $2f+1$ & \cellcolor{black!75}\color{white}$2f+1$ & \cellcolor{black!75}\color{white}$3f+1$  & $2f+1$    & $2f+1$          & $2f+1$          \\
        Determine executed requests & Completion monitor & $2f+1$ & \cellcolor{black!75}\color{white}$2f+1$ & \cellcolor{black!75}\color{white}$3f+1$ & $2f+1$    & $2f+1$          & $2f+1$          \\\hline

        \multicolumn{2}{|r||}{Total}              & $16f+8$ & $16f+8$ & $33f+13$          & $16f+8$         & $27f+13$        & $27f+13$        \\
        \multicolumn{2}{|r||}{\textbf{Byzantine fault model}} & \cellcolor{black!75}\color{white}$0$ & \cellcolor{black!75}\color{white}$16f+8$ & \cellcolor{black!75}\color{white}$33f+13$ & \cellcolor{black!75}\color{white}$5f+2$ & \cellcolor{black!75}\color{white}$5f+3$ & \cellcolor{black!75}\color{white}$7f+4$ \\\hline
        \multicolumn{2}{|l||}{Percentage of functionality to diversify (compared with baseline): $f=1$} & 0\% & 100\% & 192\% & 29\% & 33\% & 46\% \\
        \multicolumn{2}{|r||}{$f\rightarrow \infty$} & 0\% & 100\% & 206\% & 31\% & 31\% & 44\% \\
        \hline
    \end{tabular}
    \caption{Complexity comparison. Components using a Byzantine fault model are highlighted in dark. The diversification percentage compares the number of \colorbox{black!74}{\textcolor{white}{diversified components}} to the number of baseline components (e.g., $\frac{5f+2}{16f+8}$ for \minas).}
    \label{tab:eval:complexity-full}
	\vspace{-1.5mm}
\end{table*}
\shellft's modular hybrid architecture makes it possible to develop heterogeneous system implementations in which only the most exposed and/or critical components are actually diversified.
In the following, we seek to examine the ramifications of this approach with regard to complexity and programming effort.
Since both aspects are known to be difficult to quantify when it comes to actual implementations, we rely on two different methodologies for this purpose.
Our first analysis, on a more theoretical level, concentrates on the individual protocol-task functionality that needs to be diversified to improve resilience.
As a complement, our second analysis studies the code size of heterogeneous \shellft-cluster implementations we developed by applying N-version programming.

\subsection{Diversification of Protocol Functionality}
\label{sec:eval:complexity}

Our first analysis is based on the notion of a replication protocol being a composition of multiple tasks.
In \shellft protocols, each of these tasks is represented by a dedicated cluster, which allows us to use their number and sizes as a metric for complexity.
Specifically, we estimate and compare the diversification costs of different approaches by determining how often the logic of each cluster needs to be diversified.
Notice that this methodology involves simplifying assumptions: \linebreak
(1)~Although some tasks require more sophistication than others, it treats all tasks as similarly complex.
We argue that this does not pose a major problem because, as a consequence of these tasks essentially representing the atoms of a replication protocol, the overall differences are not extensive.
(2)~It does not consider general functionality such as communication, which in practice also needs to be diversified.
These parts are included in our second analysis in Section~\ref{sec:eval:n-version} which examines the actual code bases of our prototype implementations and comes to very similar conclusions, thereby confirming that the assumptions made for our first study are justified.

\vspace{-.8mm}

\headline{Baseline}
We derive the baseline for our first study from the base protocol~(see Section~\ref{sec:base-protocol}) due to its consensus mechanism being closely related to Paxos in Kirsch and Amir's system-builders variant~\cite{kirsch08paxos}, a protocol that represents the design of a typical crash-tolerant protocol in practice and applies a style of consensus that resembles the agreement process of the other analyzed protocols.
Using our metric, a common crash-tolerant system consisting of $2f+1$~replicas needs to perform all tasks handled by the 8~server-side base-protocol clusters, with every one of these tasks being implemented in each replica.
In sum, as shown in Table~\ref{tab:eval:complexity-full}, this results in \linebreak a baseline complexity equivalent to $16f+8$~micro replicas.

\vspace{-.8mm}

\headline{Traditional Hybrids}
With traditional hybrid systems such as MinBFT, XFT, and VFT~(see Section~\ref{sec:background}) consisting of monolithic replicas, diversifying them requires heterogeneous implementations of the entire replica logic (i.e.,~all base-protocol tasks), and hence leads to diversification costs of at least 100\% (Table~\ref{tab:eval:complexity-full}, ``Hybrid'' column).
Note that this number is a conservative estimate as it does not include the added logic (and thus complexity) these approaches introduce.

\vspace{-.8mm}

\headline{Full-Fledged BFT Replication}
Increasing the resilience of Byzantine fault-tolerant components by introducing heterogeneity in a Mirador-based system involves the diversification of all of its $33f+13$~micro replicas.
For $f=1$, for example, the added complexity of full-fledged BFT replication leads to overall diversification costs of $\frac{33f+13}{16f+8}=\frac{46}{24}=$~192\%.

\vspace{-.8mm}

\headline{\shellft Protocols}
\shellft's selective diversification offers the benefit of limiting diversification costs to the most exposed/critical parts: the shell clusters.
In \minas, the shell consists of only two clusters (i.e.,~front ends and executors) with a total of $5f+2$~micro replicas.
For a system tolerating a single fault in each cluster ($f=1$), this reduces the diversification efforts to only $\frac{5f+2}{16f+8}=\frac{7}{24}=$~29\% of the costs associated with traditional hybrid approaches.
Relying on larger shells, the costs for \sentry~($\frac{5f+3}{16f+8}=\frac{8}{24}=$~33\%) and \mentry~($\frac{7f+4}{16f+8}=\frac{11}{24}=$~46\%) are slightly higher, nevertheless both numbers still represent a significant improvement over traditional hybrid systems (100\%).

\vspace{-.8mm}

\headline{Impact of System Size}
To examine the relationship between diversification costs and system size we not only compute values for $f=1$, but also for $f \rightarrow \infty$.
As shown in the last line of Table~\ref{tab:eval:complexity-full}, for the three \shellft protocols the resulting numbers do not differ significantly from the ones for $f=1$, which means that the savings enabled by \shellft are largely independent of system size.
In particular, this confirms that our target systems, in which (as discussed in Section~\ref{sec:base-protocol}) $f$ is small, are already able to benefit from the approach.

\subsection{N-Version Programming}
\label{sec:eval:n-version}

\vspace{-.5mm}

In our second study, we focus on N-version programming as a means to achieve diversification.
To evaluate the impact of \shellft on implementation costs in this context, we applied the concept to all shell clusters of \minas and \sentry.
In an effort to cover a wide range of heterogeneity, we chose a variety of programming languages, using different runtime environments and even paradigms: Java (the language in which the \shellft framework itself is written), C++ (standard C++20), Go~(v1.22), as well as Elixir (v1.12 with Erlang/OTP~24), a functional programming language designed for distributed and fault-tolerant systems~\cite{elixir}.

Table~\ref{tab:eval:loc} reports the sizes of these code bases measured in lines of code~(LOC) as calculated by \texttt{cloc} (v1.90)~\cite{cloc}.
We are aware that LOC can be a somewhat imprecise unit of measurement, especially when comparing different programming languages, but it nevertheless allows us to get a good impression on the potential costs of development and maintenance.
The full-scope implementation in Java comprises the functionality for the entire replicated system and consists of around 10k\,LOC, of which about 7k\,LOC are dedicated to the protocol logic of replicas.
The difference of about 3k\,LOC includes general system functionality (e.g.,~communication, startup procedures) that is required for each replica regardless of its cluster.
When considering only the eight clusters that are part of the base protocol, this leaves around 8.9k\,LOC for the whole implementation and 5.8k\,LOC for the replicas.
In comparison, our N-version programming implementations in C++, Go and Elixir either contain two or three clusters and have a greatly reduced code size.
Specifically, they range between 2.1k\,LOC and 3.1k\,LOC, which is \mbox{21--31\%} and 23--35\% of the full-scope Java implementation and base protocol, respectively.
This difference is maintained for the replica logic and corresponds to the fact that either only two (25\%) or three (37.5\%) out of \linebreak eight clusters are diversified in another programming language.

\begin{table}
	\vspace{1.5mm}
    \centering
    \renewcommand{\arraystretch}{0.85}
    \begin{tabular}{|l|c|c|c|c|}
        \cline{2-4}
        \nocell{1} & \textbf{Scope} & \textbf{Full Code} & \textbf{Replica Logic} \\\hline
        Java     & All clusters       & $10,101$ & $7,037$    \\
        Java     & Base protocol   & $\hphantom{1}8,900$ & $5,836$    \\\hline
        C++      & Executor + front end & $\hphantom{1}2,756$ & $1,643$    \\
        Elixir   & Executor + front end & $\hphantom{1}2,126$ & $1,469$    \\
        Go       & Executor, proposer, curator & $\hphantom{1}3,081$ & $2,179$    \\\hline
    \end{tabular}
    \caption{Diverse \shellft implementations (lines of code).}
    \label{tab:eval:loc}
	\vspace{-2.2mm}
\end{table}

\subsection{Discussion}

\vspace{-.5mm}

Although applying different methodologies, both of our analyses arrive at consistent conclusions with respect to the reduction of diversification costs enabled by \shellft.
For \minas, for example, the first analysis determined the need to diversify 29\% of the functionality to improve shell resilience.
As shown by the second analysis, this closely matches the actual code sizes of our heterogeneous \minas shell-cluster implementations in C++ and Elixir.
On average, they only comprise 27\% of the size of the base-protocol code base (both in terms of the full code as well as the replica logic), and hence confirm the significant savings (more than 70\%) made possible by \linebreak \shellft compared with traditional hybridization approaches.

\section{Performance and Fault-Handling Evaluation}
\label{sec:extended-evaluation}

In this section, we present the results of experiments that compare \minas, \sentry, and \mentry against state-of-the-art protocols, evaluating both performance and the impact of various replica failures.
Notice that the goal of the performance experiments is to study our approach's impact on throughput and latency.
We do not promote \shellft as a technique for improving performance.

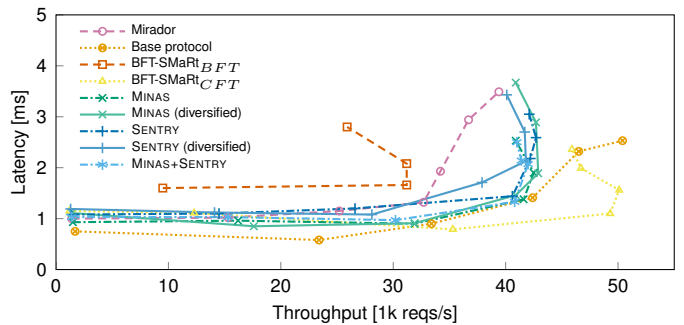
\begin{figure}[b!]
	\vspace{-2.2mm}
    \hspace{-10.8mm}%
	\newcommand{\data}[4]{
	\addplot+[dataset, mark=#1, #2,every mark/.append style={solid, semithick, scale=#3},error bars/.cd, y dir=both, y explicit] coordinates {#4}
}

\definecolor{bftsmart-cft}{HTML}{F0E442}
\definecolor{bftsmart-bft}{HTML}{D55E00}
\definecolor{mirador}{HTML}{CC79A7}
\definecolor{shellft-base}{HTML}{E69F00}
\definecolor{minas}{HTML}{009E73}
\definecolor{sentry}{HTML}{0072B2}
\definecolor{minas-sentry}{HTML}{56B4E9}

\begin{tikzpicture}[font=\scriptsize\sffamily]
	\pgfplotsset{
		compat=1.3,
		%
		every axis legend/.append style={
			legend pos=north west,
			legend cell align=left,
			legend style={font=\tiny\sffamily},
			draw=none,
			fill=none,
			inner sep=0,
			row sep=-1mm,
			xshift=.5mm,
			yshift=.5mm
		},
		%
		ybar legend/.style={
			legend image code/.code={
				\node[draw=black, minimum width=2.1mm, minimum height=1mm, fill=#1] {};
			};
		}
	}

	\tikzstyle{dataset}=[black, thick, solid];

	\begin{axis}[
		height=49.7mm,
		width=98mm,
		grid=none,
		tick align=inside,
		tick pos=left,
		major tick length=1mm,
		%
		xlabel={Throughput [1k reqs/s]},
		xlabel shift=-2pt,
		xmin=0,
		xmax=55,
		xtick={0,10,...,50},
		xticklabels={0,10,...,50},
		%
		ylabel={Latency [ms]},
		ylabel shift=-4pt,
		ymin=0,
		ymax=5,
		ytick={0,1,...,5},
		yticklabels={0,1,...,5}
	]

		\addlegendentry{Mirador}
		\data{o,densely dashed}{mirador}{0.7}{
			(1.4, 1.0)
			(15.3, 1.02)
			(25.2, 1.15)
			(32.7, 1.32)
			(34.2, 1.93)
			(36.7, 2.94)
			(39.4, 3.49)
		};

		\addlegendentry{Base protocol}
		\data{otimes,densely dotted}{shellft-base}{0.7}{
			(1.7, 0.75)
			(23.4, 0.58)
			(33.4, 0.90)
			(42.4, 1.41)
			(46.5, 2.32)
			(50.4, 2.53)
		};

		\addlegendentry{BFT-SMaRt$_{BFT}$}
		\data{square,densely dashed}{bftsmart-bft}{0.7}{
			(9.5, 1.60)
			(31.2, 1.66)
			(31.2, 2.08)
			(25.9, 2.80)
			};

		\addlegendentry{BFT-SMaRt$_{CFT}$}
		\data{triangle,densely dotted}{bftsmart-cft}{0.7}{
			(1.2, 1.15)
			(12.3, 1.11)
			(35.3, 0.79)
			(49.3, 1.10)
			(50.1, 1.56)
			(46.7, 1.99)
			(45.9, 2.36)
		};

		\addlegendentry{\minas}
		\data{x, densely dash dot}{minas}{1}{
			(1.5, 0.93)
			(16.2, 0.96)
			(31.9, 0.90)
			(41.6, 1.38)
			(42.5, 1.90)
			(40.9, 2.53)
		};

		\addlegendentry{\minas (diversified)}
		\data{x, solid}{minas!70}{1}{
			(1.4, 1.11)
			(17.6, 0.85)
			(31.9, 0.91)
			(40.9, 1.46)
			(42.9, 1.89)
			(42.7, 2.89)
			(40.9, 3.67)
		};

		\addlegendentry{\sentry}
		\data{+, densely dash dot}{sentry}{1}{
			(1.4, 1.08)
			(14.5, 1.10)
			(26.6, 1.20)
			(40.6, 1.44)
			(42.2, 2.18)
			(42.7, 2.59)
			(42.1, 3.05)
		};

		\addlegendentry{\sentry (diversified)}
		\data{+, solid}{sentry!70}{1}{
			(1.3, 1.19)
			(14.1, 1.12)
			(28.1, 1.08)
			(37.9, 1.71)
			(41.8, 2.13)
			(41.7, 2.70)
			(40.1, 3.43)
		};
		
		\addlegendentry{\mentry}
		\data{asterisk, densely dash dot}{minas-sentry}{1}{
			(1.4, 1.03)
			(15.3, 1.03)
			(30.2, 0.97)
			(40.8, 1.33)
			(42.0, 2.07)
			(41.4, 2.14)
			(41.0, 2.50)
		};
		
	\end{axis}
\end{tikzpicture}%
	\vspace{-1mm}
    \caption{Performance comparison.}
    \label{fig:evaluation-performance}
\end{figure}

\subsection{Performance}

We evaluate the performance of our \shellft protocols with multiple baselines: the base protocol, the micro-replicated BFT protocol Mirador, and the widely used \bftsmart~\cite{bessani14state} library, both in a Byzantine and a crash fault-tolerant configuration (\bftsmart$_{BFT}$ and \bftsmart$_{CFT}$).
Since \shellft primarily focuses on small systems~(see Section~\ref{sec:base-protocol}), we dimension all systems for $f\hspace{-1mm}=\hspace{-1mm}1$, meaning that \minas, \sentry and \mentry replicas are hosted on 7~machines (cf.\ Figure~\ref{fig:deployment}).
As application, we employ a key-value store and run YCSB~\cite{cooper10benchmarking} with an update-heavy workload and 1\,KB records containing fields of 100\,B values.
Each reported data point represents the average of three runs.

\headline{Throughput and Latency}
\label{sec:eval:performance}
\hspace*{-.8mm}As shown in Figure~\ref{fig:evaluation-performance}, with a maximum throughput of $\sim$42k~requests/s and a latency of $\sim$1\,ms, \linebreak the three \shellft protocols are in between the base protocol\,/\,\bftsmart$_{CFT}$ and Mirador\,/\,\bftsmart$_{BFT}$, indicating that they do not only represent an intermediate between crash tolerance and Byzantine fault tolerance in terms of complexity, but also with regard to performance.

\headline{Performance of Diversified Systems}
Comparing the pure Java implementations of \minas and \sentry with their heterogeneous counterparts comprising the diversified shell components listed in Table~\ref{tab:eval:loc}, we observe similar throughput and latency results.
This shows that a diversification of crucial parts of a replicated system as enabled by \shellft can increase resilience without noticeably impacting performance.

\subsection{Impact of Replica Failures}

\begin{figure}
	\vspace{1.8mm}
	\begin{minipage}{\columnwidth}\centering\footnotesize\textsf{\textbf{Proposer Crash and View Change}}\end{minipage}\vspace{-2mm}\\
	
    \hspace{-11mm}%

\def\myheight{26mm}
\def\mywidth{53.5mm}

\definecolor{bftsmart-cft}{HTML}{F0E442}
\definecolor{bftsmart-bft}{HTML}{D55E00}
\definecolor{mirador}{HTML}{CC79A7}
\definecolor{minas}{HTML}{009E73}
\definecolor{sentry}{HTML}{0072B2}
\definecolor{minas-sentry}{HTML}{56B4E9}

\begin{tikzpicture}[font=\scriptsize\sffamily]
	\pgfplotsset{
		compat=1.3,
		%
		every axis legend/.append style={
			legend pos=north west,
			legend cell align=left,
			draw=none,
			fill=none,
			inner sep=0,
			row sep=-1mm,
		},
		%
		ybar legend/.style={
			legend image code/.code={
				\node[draw=black, yshift=.2mm, minimum width=2.1mm, minimum height=1mm, fill=#1] {};
			};
		}
	}
	\begin{groupplot}[
		group style={
			group size=2 by 2,
			x descriptions at=edge bottom,
			horizontal sep=0.1cm,
			vertical sep=0.1cm
		},
		height=\myheight,
		width=\mywidth,
		grid=none,
		tick align=inside,
		major tick length=1mm,
		scaled y ticks = false,
		%
		xmin=-21,
		xmax=21,
		xtick={-20,-10,...,20},
		xticklabels=\empty,
		%
		ymin=0,
		ymax=17000,
		ytick={0,5000,...,15000},
		yticklabels={0,5,...,15}
	]
		

		\nextgroupplot[ytick pos=left, yticklabel pos=left]
		\addlegendentry{Mirador}
		\addplot+[mark=none, densely dotted, mirador, very thick] coordinates {
			(-21.0, 10002)
			(-20.0, 10002)
			(-19.0, 10001)
			(-18.0, 10008)
			(-17.0, 10008)
			(-16.0, 9940)
			(-15.0, 10000)
			(-14.0, 10000)
			(-13.0, 10003)
			(-12.0, 10010)
			(-11.0, 10007)
			(-10.0, 9860)
			(-9.0, 10000)
			(-8.0, 10001)
			(-7.0, 10007)
			(-6.0, 10002)
			(-5.0, 10010)
			(-4.0, 10000)
			(-3.0, 10000)
			(-2.0, 10005)
			(-1.0, 10011)
			(0.0, 10004)
			(1.0, 6066)
			(2.0, 0)
			(3.0, 0)
			(4.0, 5863)
			(5.0, 10002)
			(6.0, 9980)
			(7.0, 9979)
			(8.0, 9993)
			(9.0, 9976)
			(10.0, 9986)
			(11.0, 9991)
			(12.0, 9925)
			(13.0, 10000)
			(14.0, 9915)
			(15.0, 10000)
			(16.0, 9979)
			(17.0, 9998)
			(18.0, 9995)
			(19.0, 10004)
			(20.0, 9957)
			(21.0, 9990)
		};

		\nextgroupplot[ytick pos=right, yticklabel pos=right]
		\addlegendentry{\mentry}
		\addplot+[mark=none, densely dotted, minas-sentry, very thick] coordinates {
			(-21.0, 10016)
			(-20.0, 9994)
			(-19.0, 10009)
			(-18.0, 10001)
			(-17.0, 10002)
			(-16.0, 9986)
			(-15.0, 10001)
			(-14.0, 10010)
			(-13.0, 10000)
			(-12.0, 10000)
			(-11.0, 10005)
			(-10.0, 9998)
			(-9.0, 9886)
			(-8.0, 10001)
			(-7.0, 10005)
			(-6.0, 10000)
			(-5.0, 10003)
			(-4.0, 10005)
			(-3.0, 10006)
			(-2.0, 10002)
			(-1.0, 10004)
			(0.0, 9081)
			(1.0, 0)
			(2.0, 0)
			(3.0, 5627)
			(4.0, 9994)
			(5.0, 10003)
			(6.0, 9994)
			(7.0, 10007)
			(8.0, 9999)
			(9.0, 9997)
			(10.0, 9995)
			(11.0, 10000)
			(12.0, 10001)
			(13.0, 10007)
			(14.0, 9994)
			(15.0, 9997)
			(16.0, 10009)
			(17.0, 10000)
			(18.0, 10000)
			(19.0, 10001)
			(20.0, 9999)
			(21.0, 9964)
		};

		\nextgroupplot[
			ytick pos=left,
			yticklabel pos=left,
			ylabel={\hspace*{-2.6mm}Throughput [1k req/s]},
			ylabel shift=-4pt,
			ylabel style={at=(ticklabel cs:1.1)},
		]
		\addlegendentry{\minas}
		\addplot+[mark=none, densely dotted, minas, very thick] coordinates {
			(-21.0, 9938)
			(-20.0, 10000)
			(-19.0, 10004)
			(-18.0, 9996)
			(-17.0, 10015)
			(-16.0, 10004)
			(-15.0, 9981)
			(-14.0, 10000)
			(-13.0, 10008)
			(-12.0, 10008)
			(-11.0, 9995)
			(-10.0, 10002)
			(-9.0, 10000)
			(-8.0, 10001)
			(-7.0, 10000)
			(-6.0, 10015)
			(-5.0, 10004)
			(-4.0, 9995)
			(-3.0, 10006)
			(-2.0, 10000)
			(-1.0, 10000)
			(0.0, 10013)
			(1.0, 9947)
			(2.0, 0)
			(3.0, 4142)
			(4.0, 10009)
			(5.0, 10008)
			(6.0, 10000)
			(7.0, 10000)
			(8.0, 10000)
			(9.0, 10001)
			(10.0, 10013)
			(11.0, 10003)
			(12.0, 10003)
			(13.0, 10000)
			(14.0, 10000)
			(15.0, 10000)
			(16.0, 10008)
			(17.0, 10012)
			(18.0, 10000)
			(19.0, 10000)
			(20.0, 10000)
			(21.0, 10001)
		};

		\nextgroupplot[ytick pos=right, yticklabel pos=right]
		\addlegendentry{\sentry}
		\addplot+[mark=none, densely dotted, sentry, very thick] coordinates {
			(-21.0, 10000)
			(-20.0, 10004)
			(-19.0, 10003)
			(-18.0, 10002)
			(-17.0, 10001)
			(-16.0, 10008)
			(-15.0, 9998)
			(-14.0, 9996)
			(-13.0, 10006)
			(-12.0, 10000)
			(-11.0, 10003)
			(-10.0, 10008)
			(-9.0, 10001)
			(-8.0, 10002)
			(-7.0, 10005)
			(-6.0, 10003)
			(-5.0, 10001)
			(-4.0, 10007)
			(-3.0, 10002)
			(-2.0, 10002)
			(-1.0, 10001)
			(0.0, 2459)
			(1.0, 0)
			(2.0, 2191)
			(3.0, 9908)
			(4.0, 10005)
			(5.0, 9918)
			(6.0, 10013)
			(7.0, 9992)
			(8.0, 10014)
			(9.0, 10000)
			(10.0, 10001)
			(11.0, 10000)
			(12.0, 10002)
			(13.0, 9998)
			(14.0, 10000)
			(15.0, 9979)
			(16.0, 10012)
			(17.0, 9999)
			(18.0, 10000)
			(19.0, 9990)
			(20.0, 10004)
			(21.0, 10008)
		};

	\end{groupplot}

\end{tikzpicture}

	
	\begin{minipage}{\columnwidth}\centering\footnotesize\textsf{\textbf{Byzantine-faulty Proposer}}\end{minipage}\vspace{-2mm}\\
    
	\hspace*{-12.25mm}%

\def\myheight{26mm}
\def\mywidth{53.5mm}

\definecolor{bftsmart-cft}{HTML}{F0E442}
\definecolor{bftsmart-bft}{HTML}{D55E00}
\definecolor{mirador}{HTML}{CC79A7}
\definecolor{minas}{HTML}{009E73}
\definecolor{sentry}{HTML}{0072B2}
\definecolor{minas-sentry}{HTML}{56B4E9}
\definecolor{shellft-base}{HTML}{E69F00}

\begin{tikzpicture}[font=\scriptsize\sffamily]
	\pgfplotsset{
		compat=1.3,
		%
		every axis legend/.append style={
			legend pos=north west,
			legend cell align=left,
			draw=none,
			fill=none,
			inner sep=0,
			row sep=-1mm,
		},
		%
		ybar legend/.style={
			legend image code/.code={
				\node[draw=black, yshift=.2mm, minimum width=2.1mm, minimum height=1mm, fill=#1] {};
			};
		}
	}

	\begin{groupplot}[
        group style={
            group size=2 by 2,
            x descriptions at=edge bottom,
            horizontal sep=0.1cm,
            vertical sep=0.1cm
        },
		height=\myheight,
		width=\mywidth,
		grid=none,
		tick align=inside,
		major tick length=1mm,
		scaled y ticks = false,
		%
		xmin=-21,
		xmax=21,
		xtick={-20,-10,...,20},
		xticklabels=\empty,
		%
		ymin=0,
		ymax=17000,
		ytick={0,5000,...,15000},
		yticklabels={0,5,...,15}
	]
		
        \nextgroupplot[ytick pos=left, yticklabel pos=left]
        \addlegendentry{Mirador}
        \addplot+[mark=none, densely dotted, mirador, very thick] coordinates {
            (-21.0, 10010)
            (-20.0, 9830)
            (-19.0, 9981)
            (-18.0, 10011)
            (-17.0, 9972)
            (-16.0, 10005)
            (-15.0, 10003)
            (-14.0, 9868)
            (-13.0, 10018)
            (-12.0, 10007)
            (-11.0, 9996)
            (-10.0, 10010)
            (-9.0, 9989)
            (-8.0, 9993)
            (-7.0, 10001)
            (-6.0, 10004)
            (-5.0, 9997)
            (-4.0, 10005)
            (-3.0, 10000)
            (-2.0, 10004)
            (-1.0, 10006)
            (0.0, 10004)
            (1.0, 984)
            (2.0, 0)
            (3.0, 302)
            (4.0, 9735)
            (5.0, 9998)
            (6.0, 9995)
            (7.0, 10007)
            (8.0, 10005)
            (9.0, 9971)
            (10.0, 10011)
            (11.0, 10005)
            (12.0, 10002)
            (13.0, 10000)
            (14.0, 9997)
            (15.0, 9999)
            (16.0, 10015)
            (17.0, 10005)
            (18.0, 9999)
            (19.0, 10001)
            (20.0, 10000)
            (21.0, 10000)
        };

        \nextgroupplot[ytick pos=right, yticklabel pos=right]
        \addlegendentry{Base protocol}
        \addplot+[mark=none, densely dotted, shellft-base, very thick] coordinates {
            (-21.0, 10002)
            (-20.0, 9958)
            (-19.0, 9985)
            (-18.0, 10032)
            (-17.0, 10000)
            (-16.0, 10000)
            (-15.0, 10000)
            (-14.0, 10012)
            (-13.0, 10008)
            (-12.0, 9985)
            (-11.0, 10015)
            (-10.0, 10000)
            (-9.0, 9960)
            (-8.0, 9942)
            (-7.0, 9824)
            (-6.0, 9799)
            (-5.0, 10008)
            (-4.0, 10000)
            (-3.0, 10000)
            (-2.0, 10002)
            (-1.0, 9924)
            (0.0, 9692)
            (1.0, 2334)
            (2.0, 0)
            (21.0, 0)
        };

        \nextgroupplot[
            ytick pos=left,
            yticklabel pos=left,
            ylabel={\hspace*{-2.6mm}Throughput [1k req/s]},
		    ylabel shift=-4pt,
            ylabel style={at=(ticklabel cs:1.1)},
        ]
        \addlegendentry{\minas}
        \addplot+[mark=none, densely dotted, minas, very thick] coordinates {
            (-21.0, 10012)
            (-20.0, 10000)
            (-19.0, 10000)
            (-18.0, 10002)
            (-17.0, 10000)
            (-16.0, 10004)
            (-15.0, 9891)
            (-14.0, 10001)
            (-13.0, 10002)
            (-12.0, 9997)
            (-11.0, 10018)
            (-10.0, 10002)
            (-9.0, 10000)
            (-8.0, 10000)
            (-7.0, 10000)
            (-6.0, 10000)
            (-5.0, 10013)
            (-4.0, 10007)
            (-3.0, 10000)
            (-2.0, 10000)
            (-1.0, 10000)
            (0.0, 10000)
            (1.0, 6132)
            (2.0, 0)
            (21.0, 0)
        };

        \nextgroupplot[ytick pos=right, yticklabel pos=right]
        \addlegendentry{\sentry}
        \addplot+[mark=none, densely dotted, sentry, very thick] coordinates {
            (-21.0, 9955)
            (-20.0, 10002)
            (-19.0, 10013)
            (-18.0, 9998)
            (-17.0, 10007)
            (-16.0, 10002)
            (-15.0, 10000)
            (-14.0, 9999)
            (-13.0, 10001)
            (-12.0, 10005)
            (-11.0, 10007)
            (-10.0, 10006)
            (-9.0, 10001)
            (-8.0, 10001)
            (-7.0, 9987)
            (-6.0, 10006)
            (-5.0, 10006)
            (-4.0, 10006)
            (-3.0, 10001)
            (-2.0, 10000)
            (-1.0, 10010)
            (0.0, 5256)
            (1.0, 0)
            (2.0, 0)
            (3.0, 5588)
            (4.0, 10000)
            (5.0, 10004)
            (6.0, 9998)
            (7.0, 10009)
            (8.0, 10008)
            (9.0, 9997)
            (10.0, 10006)
            (11.0, 10000)
            (12.0, 10002)
            (13.0, 10004)
            (14.0, 10003)
            (15.0, 10000)
            (16.0, 10008)
            (17.0, 10010)
            (18.0, 9999)
            (19.0, 10002)
            (20.0, 10004)
            (21.0, 10003)
        };

	\end{groupplot}

\end{tikzpicture}


	\begin{minipage}{\columnwidth}\centering\footnotesize\textsf{\textbf{Byzantine-faulty Executor}}\end{minipage}\vspace{-2mm}\\
	
    \hspace{-12.25mm}%

\def\myheight{26mm}
\def\mywidth{53.5mm}

\definecolor{bftsmart-cft}{HTML}{F0E442}
\definecolor{bftsmart-bft}{HTML}{D55E00}
\definecolor{mirador}{HTML}{CC79A7}
\definecolor{minas}{HTML}{009E73}
\definecolor{sentry}{HTML}{0072B2}
\definecolor{minas-sentry}{HTML}{56B4E9}
\definecolor{shellft-base}{HTML}{E69F00}

\begin{tikzpicture}[font=\scriptsize\sffamily]
	\pgfplotsset{
		compat=1.3,
		%
		every axis legend/.append style={
			legend pos=north west,
			legend cell align=left,
			draw=none,
			fill=none,
			inner sep=0,
			row sep=-1mm,
		},
		%
		ybar legend/.style={
			legend image code/.code={
				\node[draw=black, yshift=.2mm, minimum width=2.1mm, minimum height=1mm, fill=#1] {};
			};
		}
	}

	\begin{groupplot}[
        group style={
            group size=2 by 2,
            x descriptions at=edge bottom,
            horizontal sep=0.1cm,
            vertical sep=0.1cm
        },
		height=\myheight,
		width=\mywidth,
		grid=none,
		tick align=inside,
		tick pos=bottom,
		major tick length=1mm,
		scaled y ticks = false,
		%
		xmin=-21,
		xmax=21,
		xtick={-20,-10,...,20},
		xticklabels={\hspace*{1mm}-20,-10,0,10,20},
		%
		ymin=0,
		ymax=17000,
		ytick={0,5000,...,15000},
		yticklabels={0,5,...,15}
	]
		
        \nextgroupplot[ytick pos=left, yticklabel pos=left]
        \addlegendentry{Mirador}
        \addplot+[mark=none, densely dotted, mirador, very thick] coordinates {
			(-21.0, 10000)
			(-20.0, 10000)
			(-19.0, 10004)
			(-18.0, 10008)
			(-17.0, 10006)
			(-16.0, 10002)
			(-15.0, 10000)
			(-14.0, 10000)
			(-13.0, 10000)
			(-12.0, 10003)
			(-11.0, 10006)
			(-10.0, 10011)
			(-9.0, 10000)
			(-8.0, 10000)
			(-7.0, 10000)
			(-6.0, 10000)
			(-5.0, 9992)
			(-4.0, 10007)
			(-3.0, 10009)
			(-2.0, 10003)
			(-1.0, 9980)
			(0.0, 9779)
			(1.0, 9332)
			(2.0, 9919)
			(3.0, 9934)
			(4.0, 9903)
			(5.0, 9917)
			(6.0, 9923)
			(7.0, 9940)
			(8.0, 9870)
			(9.0, 9934)
			(10.0, 9918)
			(11.0, 9833)
			(12.0, 9941)
			(13.0, 9908)
			(14.0, 9946)
			(15.0, 9865)
			(16.0, 9936)
			(17.0, 9934)
			(18.0, 9937)
			(19.0, 9943)
			(20.0, 9911)
			(21.0, 9948)
        };

        \nextgroupplot[ytick pos=right, yticklabel pos=right]
        \addlegendentry{Base protocol}
        \addplot+[mark=none, densely dotted, shellft-base, very thick] coordinates {
			(-21.0, 10000)
			(-20.0, 10000)
			(-19.0, 10000)
			(-18.0, 10000)
			(-17.0, 10008)
			(-16.0, 10012)
			(-15.0, 10000)
			(-14.0, 10000)
			(-13.0, 10000)
			(-12.0, 10003)
			(-11.0, 10006)
			(-10.0, 9991)
			(-9.0, 10020)
			(-8.0, 10000)
			(-7.0, 10000)
			(-6.0, 10008)
			(-5.0, 9992)
			(-4.0, 10006)
			(-3.0, 9993)
			(-2.0, 9885)
			(-1.0, 9922)
			(0.0, 7837)
			(1.0, 0)
			(21.0, 0)
        };

        \nextgroupplot[
            xlabel={Time [s]},
			xlabel shift=-4pt,
            xlabel style={at=(ticklabel cs:1.1)},
            ytick pos=left,
            yticklabel pos=left,
            ylabel={\hspace*{-2.6mm}Throughput [1k req/s]},
			ylabel shift=-4pt,
            ylabel style={at=(ticklabel cs:1.1)},
        ]
        \addlegendentry{\minas}
        \addplot+[mark=none, densely dotted, minas, very thick] coordinates {
			(-21.0, 10008)
			(-20.0, 10007)
			(-19.0, 10000)
			(-18.0, 10000)
			(-17.0, 10000)
			(-16.0, 10000)
			(-15.0, 10006)
			(-14.0, 10002)
			(-13.0, 10012)
			(-12.0, 10000)
			(-11.0, 10000)
			(-10.0, 10000)
			(-9.0, 9932)
			(-8.0, 10007)
			(-7.0, 10002)
			(-6.0, 9999)
			(-5.0, 10006)
			(-4.0, 10001)
			(-3.0, 10008)
			(-2.0, 10000)
			(-1.0, 10003)
			(0.0, 10006)
			(1.0, 9997)
			(2.0, 10005)
			(3.0, 10000)
			(4.0, 10009)
			(5.0, 10005)
			(6.0, 9901)
			(7.0, 9938)
			(8.0, 9930)
			(9.0, 10000)
			(10.0, 10000)
			(11.0, 10013)
			(12.0, 10003)
			(13.0, 10004)
			(14.0, 10000)
			(15.0, 10000)
			(16.0, 9933)
			(17.0, 10005)
			(18.0, 9995)
			(19.0, 10011)
			(20.0, 10002)
			(21.0, 9946)
        };

        \nextgroupplot[ytick pos=right, yticklabel pos=right]
        \addlegendentry{\sentry}
        \addplot+[mark=none, densely dotted, sentry, very thick] coordinates {
			(-21.0, 10005)
			(-20.0, 10001)
			(-19.0, 9839)
			(-18.0, 9961)
			(-17.0, 10000)
			(-16.0, 10001)
			(-15.0, 10000)
			(-14.0, 10019)
			(-13.0, 9997)
			(-12.0, 10003)
			(-11.0, 10000)
			(-10.0, 10001)
			(-9.0, 9999)
			(-8.0, 10007)
			(-7.0, 10000)
			(-6.0, 10003)
			(-5.0, 10010)
			(-4.0, 10001)
			(-3.0, 9923)
			(-2.0, 10016)
			(-1.0, 9998)
			(0.0, 10002)
			(1.0, 10000)
			(2.0, 10000)
			(3.0, 10017)
			(4.0, 10003)
			(5.0, 9996)
			(6.0, 10004)
			(7.0, 10000)
			(8.0, 10000)
			(9.0, 10000)
			(10.0, 10010)
			(11.0, 10010)
			(12.0, 10000)
			(13.0, 9980)
			(14.0, 10000)
			(15.0, 10000)
			(16.0, 10008)
			(17.0, 10012)
			(18.0, 10000)
			(19.0, 10000)
			(20.0, 10000)
			(21.0, 10000)
        };

	\end{groupplot}

\end{tikzpicture}

    \vspace*{-1.7em}%
	
    \caption{Impact of different fault scenarios.}
    \label{fig:evaluation-faults}
	\vspace{-.6mm}
\end{figure}
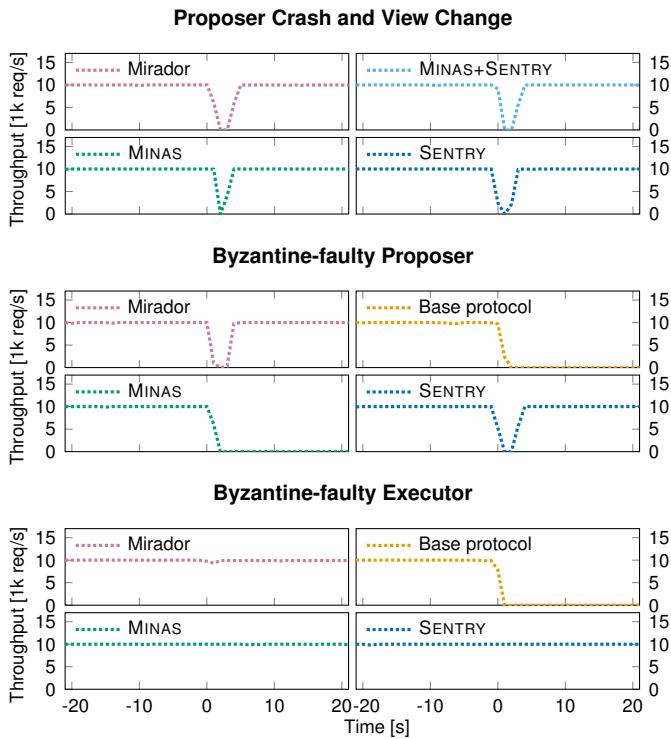

The final part of our evaluation investigates the impact of replica failures, which is why we subject the systems to both replica crashes as well as Byzantine behavior~(see Figure~\ref{fig:evaluation-faults}).
The view-change timeout in these experiments is set to 1\,s.

\headline{Leader Crash}
As the consequence of a crash of the current proposer, all three \shellft protocols experience the expected downtime of 1--2\,s and resume a stable performance after the successful view change. The view change in \minas is completed slightly faster due to comprising fewer phases.

\headline{Byzantine Failures}
To study the impact of arbitrary faults on two key clusters (i.e., proposer and executor), we employ the full extent of their potential Byzantine behavior, ranging from equivocation to forging responses and checkpoints.
Again, all evaluated protocols demonstrate the behavior expected from their respective fault model.
While (Byzantine) equivocation of the proposer in \minas and the base protocol leads to diverging replicas, Mirador and \sentry (as well as \mentry) tolerate the fault by issuing a view change.
In contrast, with executors being part of their shell, all \shellft protocols tolerate a Byzantine-faulty executor.
Unlike the base protocol, none \linebreak of the \shellft protocols experiences any notable impact.

\section{Related Work}
\label{sec:related}

Having already discussed a large body of related work in the area of hybridization in Section~\ref{sec:background}, in the following we focus on additional aspects with relevance to \shellft.

\headline{Modularized Replication Architectures}
Micro replication so far has only been investigated in the context of improving debuggability~\cite{distler2023micro}.
With \shellft, we are the first to harness its properties for the design of flexible hybrid architectures.

Whittaker et al.~\cite{whittaker21scaling} presented a systematic approach to eliminate performance bottlenecks by compartmentalizing the affected replication-protocol parts into multiple components.
Concentrating on efficiency and scalability, this method is orthogonal to our goal of combining different fault models within the same system.
However, being a general technique, compartmentalization could be used to improve performance in \shellft protocols.
The same is true for mechanisms improving the communication flow between replicas~\mbox{\cite{charapko2021pigpaxos,batista23flexcast}}.

\vspace{-.5mm}

\headline{N-Versioning of Microservices}
Espinoza et al.~\cite{espinoza2022back} showed that heterogeneous implementations are an effective means to increase robustness in microservice architectures, especially when applied to components handling user data~(e.g., input sanitizers, which in \shellft can be implemented in the front ends).
Unlike \shellft, their approach does not consider replication protocols and introduces non-replicated (and hence non-fault-tolerant) proxies for each diversified microservice.

\vspace{-.5mm}

\headline{Physical Separation of Replicas}
Placing replicas at significant geographic distances from each other, geo-replicated systems such as GeoPaxos~\cite{coelho18geographic}, GeoPaxos+~\cite{coelho21geopaxosplus}, or ATLAS~\cite{enes2020state} make it highly unlikely that a single root cause leads to the failure of multiple replicas.
Like these systems, a \shellft deployment relies on a physical separation between replicas providing the same functionality~(i.e.,~\shellft replicas belonging to the same cluster),
but in addition it also isolates shell replicas from filters/cores.
Studying the benefits and implications of implementing our two-level separation in geo-re\-pli\-ca\-ted settings is an interesting direction for future work.

\vspace{-.5mm}

\headline{Adaptation to Threats}
Targeting scenarios in which the strength of an adversary evolves over time, Silva et al.~\cite{silva21threat} developed a BFT system that is able to dynamically adapt both the number of replicas and its resilience threshold at runtime.
Integrating this concept with \shellft protocols (at the \linebreak cluster level) would further decrease their resource footprint.

\vspace{-.5mm}

\headline{Hardening}
Correia et al.~\cite{correia12practical} proposed to extend \mbox{crash-tole}\-rant systems with integrity checks to detect arbitrary state corruption and avoid error propagation through erroneous messages.
Behrens et al.~\cite{behrens14hardpaxos} showed that for Paxos these kinds of checks can be limited to a small part of the protocol logic.
Compared with the diversification of replicas, such hardening techniques incur less development costs but, on the other hand, \linebreak are only able to cope with a subset of Byzantine faults.
\section{Conclusion}

\shellft's selective hybridization offers an unprecedented degree of flexibility when it comes to tailoring the resilience of replicated systems to specific use cases.
At the same time, \shellft is able to decrease diversification costs by more than 70\% compared with traditional hybridization approaches.

\bibliographystyle{IEEEtran}
\bibliography{bib/extended}

\appendices

\section{Proofs}
\label{sec:proofs}

\vspace{-.3mm}

As explained in Section~\ref{sec:shellf:transformation}, \shellft's protocol transformation process exchanges architectural patterns from the crash-tolerant domain with their corresponding patterns from the Byzantine fault-tolerant domain while maintaining safety and liveness guarantees.
In the following, we present the associated proofs for the base protocol's two main patterns.

\subsection{Reliable Distribution Pattern (Figure~\ref{fig:transformation-reliable})}
\label{sec:proofs-rdp}

\newtheorem{propertyrdpapp}{Property}
\renewcommand{\thepropertyrdpapp}{RDP.\arabic{propertyrdpapp}}

\begin{propertyrdpapp}
	\label{the:app:reliable-distribution-1}
	If a correct sink~$s_1$ accepts a value~$v$ and another correct sink~$s_2$ accepts a value~$v'$, then $v = v'$.
\end{propertyrdpapp}

\begin{proof}
	For crash faults, this property is a direct consequence of the fact that the source does not make any conflicting proposals to different sinks.
	In contrast, a Byzantine source may propose different values to different witnesses.
	However, due to $2f+1$ matching opinions from different witnesses being required for a correct sink to accept a value, having a total of $3f+1$ witnesses guarantees that only at most one value is able to reach the necessary quorum to be accepted.
\end{proof}

\begin{propertyrdpapp}
	\label{the:app:reliable-distribution-2}
	If the source is correct and proposes~$v$, all correct sinks eventually accept~$v$, even if $f$ witnesses are faulty.
\end{propertyrdpapp}

\begin{proof}
	If the source is correct, then its proposal will eventually arrive at all correct designated receivers (possibly after retransmissions), which in the base version of the pattern are the sinks.
	In the transformed version, eventually $2f+1$~correct witnesses receive the proposal, which is sufficient for the value to be eventually accepted by all correct sinks.
\end{proof}

\subsection{Relay Pattern (Figure~\ref{fig:transformation-relay})}
\label{sec:proofs-rp}

\newtheorem{propertyrpapp}{Property}
\renewcommand{\thepropertyrpapp}{RP.\arabic{propertyrpapp}}

\begin{propertyrpapp}
	\label{the:app:relay-1}
	If a correct sink accepts a value~$v$, then $v$ was proposed by a correct source.
\end{propertyrpapp}

\begin{proof}
	For crash faults, all sources propose the same value (or none), and relays accept a value based on matching inputs from $f+1$~sources, of which at least one is correct.
	In the transformed version, $f$~Byzantine sources may propose diverging values, however these are not sufficient to pass the threshold of $t_s = 2f+1$~matching source inputs required for a correct relay to accept a value.
	Being part of the filter, relays only forward accepted values to both sinks and other relays.
	Thus, only \linebreak the value of a correct source is able to reach a correct sink.
\end{proof}

\begin{propertyrpapp}
	\label{the:app:relay-2}
	If a correct sink accepts a value~$v$, then all correct sinks eventually accept~$v$, even if $f$ relays are faulty.
\end{propertyrpapp}

\begin{proof}
	A correct sink accepts a value~$v$ after obtaining matching opinions from $f+1$~relays, at least one of which is correct.
	Even if $f$~relays fail (in this context it does not matter if this is by crashing or in a Byzantine way), the cluster-internal propagation ensures that in such case eventually $f+1$~correct relays receive~$v$.
	This in turn guarantees that all correct sinks eventually obtain $f+1$~matching relay opinions for~$v$.
\end{proof}

\section{Base-Protocol Specifics}
\label{sec:specification}

This section provides details on the system model of our base protocol and the properties it guarantees under these conditions.
In addition, we present its complete specification.

\subsection{System Model and Guarantees}

In order to be widely applicable, the base protocol was designed using common assumptions made for crash-tolerant replication protocols in practice~\cite{lamport98part,ongaro14search}.
Among other things, this includes replicas being connected via an unreliable network that potentially delays, drops, or reorders messages; however, if a correct sender repeatedly transmits the same message to the same correct receiver, then the message will eventually arrive at its destination.

With regard to replicas, the base protocol addresses crash-stop failures; that is, replicas are either correct (in which case they behave according to specification) or crashed (in which case they cease protocol execution and do no longer communicate with other replicas or clients).
The protocol is able to tolerate up to $f$~of such replica failures in each of its eight clusters.
For this purpose, depending on the particular task they fulfill, clusters consist of $f+1$~(proposer cluster) or $2f+1$~(all other clusters)~micro replicas.

As is common for state-machine replication, the base protocol provides two key properties: safety and liveness.
While safety is ensured at all times, liveness depends on partial synchrony~\cite{dwork88consensus}, which in a nutshell means that there must be sufficiently long synchronous periods with an upper bound on processing and network delays.

\renewcommand{\proofname}{Proof Sketch}
\newtheorem{propertybp}{Property}
\renewcommand{\thepropertybp}{BP.\arabic{propertybp} (Safety)}

\begin{propertybp}
	\label{prop:bp-safety}
	If the two command sequences $\langle x_1, x_2,..., x_j \rangle$ and $\langle x^\prime_1, x^\prime_2,..., x_{j^\prime} \rangle$ are committed by two correct executor replicas, then $x_i = x^\prime_i$ for all $i \leq min(j, j^\prime)$.
\end{propertybp}

\begin{proof}
	Together, the three main base-protocol clusters (i.e., proposer, committer, and executor) implement a Paxos-style agreement process~\cite{kirsch08paxos} that represents a uniform, reliable, totally ordered multicast with the executors acting as receivers, and for this reason ensures Property~\ref{prop:bp-safety}.
\end{proof}

\renewcommand{\thepropertybp}{BP.\arabic{propertybp} (Liveness)}

\begin{propertybp}
	If a client-issued command~$x$ is received by at least one correct front-end replica, then all correct executor replicas will eventually process command~$x$.
\end{propertybp}

\begin{proof}
	With front ends exchanging new commands among each other~(see Appendix~\ref{spec:fre}, Lines~\ref{spec:fre:fetch-start}--\ref{spec:fre:fetch-end}), it is ensured that if a command~$x$ arrives at a correct front-end replica, then eventually all correct front-end replicas will obtain the command; this is true even if the client that issued the command fails in the meantime.
	Since at least $f+1$~front ends are correct, command~$x$ will eventually be reflected in the command-progress information they report to controllers, and consequently lead to all correct controllers updating their progress targets accordingly~(see Appendix~\ref{spec:ctr}, Lines~\mbox{\ref{spec:ctr:target-start}--\ref{spec:ctr:target-end}}).
	
	Due to controllers relying on these targets to monitor the agreement process, there are two possible scenarios:
	(1)~At least one executor commits command~$x$ before the controller cluster triggers a view change; in this case, the safety and liveness properties of the Paxos-style consensus guarantee that eventually all correct executors commit command~$x$.
	(2)~The controllers announce a new view before command~$x$ is agreed on; in this case, the consensus process is retried with a different proposer replica acting as current leader.
	Either way, with at most $f$ of the $f+1$~proposers being faulty and controllers repeatedly increasing the view-change timeout while no progress is made~(see Appendix~\ref{spec:ctr}, Line~\ref{spec:ctr:back-off}), under partial synchrony at some point (possibly after multiple view changes) all correct executors commit and process command~$x$.
	
	As control loops implement the relay pattern (and hence provide Properties~\ref{the:app:relay-1} and~\ref{the:app:relay-2}), it is ensured that all correct replicas eventually shift their windows forward.
	This way, the process above is able to continue for further commands.
\end{proof}

\def\possible{$\top$}
\def\impossible{$\bot$}
\def\emptyslot{$\clubsuit$}

\newpage%

\subsection{Data Structures}

\vspace{-2mm}

\begin{lstlisting}[aboveskip=1mm]
/* If-then-else helper function%\,%*/
%\textsc{Any}% %\lstbtt{ite}%(%\textsc{Boolean}% v; %\textsc{Any}% a, %\textsc{Any}% b) {
	%\pc{If }%(v == true) return a;
	return b;
}%\vspace{1.5mm}\hrule\vspace{1.5mm}%
interface %\textsc{Set}%<%\textsc{V}%> {
	/* Operations%\,%*/
	void %\lstbtt{add}%(%\textsc{V}% value);
	void %\lstbtt{delete}%(%\textsc{V}% value);
	%\textsc{Number}% size();
}

interface %\textsc{Map}%<%\textsc{K}%, %\textsc{V}%> {
	/* State%\,%*/
	%\textsc{Set}%<%\textsc{K}%> %\pc{$keys$}%;
	%\textsc{Set}%<%\textsc{V}%> %\pc{$values$}%;
	
	/* Operations%\,%*/
	void %\lstbtt{put}%(%\textsc{K}% key, %\textsc{V}% value);%\hfill%/* Key accessed via %\lsttt{[]}% operator%\,%*/
	%\textsc{V}% %\lstbtt{get}%(%\textsc{K}% key);%\hfill%/* Accessed via %\lsttt{[]}% operator%\,%*/
	void %\lstbtt{delete}%(%\textsc{K}% key);
	%\textsc{Number}% size();
}%\vspace{1.5mm}\hrule\vspace{1.5mm}%
class %\textsc{Range}%<%\textsc{N}\pc{ is a }\textsc{Number}%> {
	/* State%\,%*/
	%\textsc{N}% %\pc{$from$}%;
	%\textsc{N}% %\pc{$count$}%;

	/* Constructor%\,%*/
	%\textsc{Range}%(%\textsc{N}% from, %\textsc{N}% count) {
		%\pc{$from$}% := from;
		%\pc{$count$}% := count;
	}
}

class %\textsc{Sequence}%<%\textsc{N}\pc{ is a }\textsc{Number}%, %\textsc{V}%> {
	/* State%\,%*/
	%\textsc{N}% %\pc{$capacity$}%;
	%\textsc{N}% %\pc{$min$}%;
	%\textsc{N}% %\pc{$max$}%;
	%\textsc{N}% %\pc{$pos$}%;
	%\textsc{V}%[] %\pc{$values$}%;
	
	/* Constructor%\,%*/
	%\textsc{Sequence}%(%\textsc{N}% min, %\textsc{N}% max) {
		%\pc{$capacity$}% := max - min;
		%\pc{$min$}% := min;
		%\pc{$max$}% := max;
		%\pc{$pos$}% := min;
		%\pc{$values$}% := %\textsc{V}%[%\pc{$capacity$}%];
	}
	
	/* Operation%\,%*/
	void %\lstbtt{put}%(%\textsc{N}% index, %\textsc{V}% value) {%\hfill%/* Index accessed via %\lsttt{[]}% operator%\,%*/
		/* Check state and input%\,%*/
		%\pc{If }%(%\pc{$pos$}% == %\pc{$max$}%) return;
		%\pc{If }%(%\pc{$pos$}% != index) return;
		
		/* Update state%\,%*/
		%\pc{$values$}%[index - %\pc{$min$}%] := value;
		%\pc{$pos$}% := index + 1;
	}
	
	%\textsc{V}% %\lstbtt{get}%(%\textsc{N}% index) {%\hfill%/* Index accessed via %\lsttt{[]}% operator%\,%*/
		%\pc{If }%(index < %\pc{$min$}%) return %\pc{nil}%;
		%\pc{If }%(index >= %\pc{$pos$}%) return %\pc{nil}%;
		return %\pc{$values$}%[index - %\pc{$min$}%];
	}
}

typedef %\textsc{Ranges}%<%\textsc{I}\pc{ is an }\textsc{ID}%, %\textsc{N}\pc{ is a }\textsc{Number}%>: %\textsc{Map}%<%\textsc{I}%, %\textsc{Range}%<%\textsc{N}%>>;
typedef %\textsc{Sequences}%<%\textsc{I}\pc{ is an }\textsc{ID}%, %\textsc{N}\pc{ is a }\textsc{Number}%, %\textsc{V}%>:%\\%%\textsc{Map}%<%\textsc{I}%, %\textsc{Sequence}
%<%\textsc{N}%, %\textsc{V}%>>;%\vspace{1.5mm}\hrule\vspace{1.5mm}\pagebreak%
class %\textsc{Window}%<%\textsc{N}\pc{ is a }\textsc{Number}%,%\,\textsc{V}%> extends %\textsc{Sequence}%<%\textsc{N}%,%\,\textsc{V}%> {
	/* Operations%\,%*/
	void %\lstbtt{fill}%(%\textsc{N}% to, %\textsc{V}% value) {
		%\pc{For each \textsc{N} }%index%\pc{ in }%[%\pc{$pos$}%, min(to, %\pc{$max$}%)]: this[index] := value;
	}
	
	void %\lstbtt{move}%(%\textsc{N}% min) {
		/* Only move forward%\,%*/
		%\pc{If }%(min <= %\pc{$min$}%) return;
		
		/* Determine state%\,%*/
		%\textsc{V}%[] values := %\textsc{V}%[%\pc{$capacity$}%];
		%\pc{For each \textsc{N} }%index%\pc{ in }%[min, min + %\pc{$capacity$}%] {
			values[index - min] := this[index];
		}
		
		/* Update state%\,%*/
		%\pc{$min$}% := min;
		%\pc{$max$}% := min + %\pc{$capacity$}%;
		%\pc{$pos$}% := max(%\pc{$pos$}%, min);
		%\pc{$values$}% := values;
	}
	
	void %\lstbtt{move}%(%\textsc{N}% min, %\textsc{V}% value) {
		move(min);
		fill(%\pc{$max$}%, value);
	}
	
	void %\lstbtt{clear}%(%\textsc{N}% from) {
		%\textsc{N}% start := max(from, %\pc{$min$}%);
		%\pc{For each \textsc{N} }%index%\pc{ in }%[start, %\pc{$pos$}%]: %\pc{$values$}%[index - %\pc{$min$}%] := %\pc{nil}%;
		%\pc{$pos$}% := start;
	}
	
	void %\lstbtt{reset}%() {
		clear(%\pc{$min$}%);
	}
	
	void %\lstbtt{sync}%(%\textsc{Window}%<%\textsc{N}%, %\textsc{*}%> window) {
		move(window.%\pc{$min$}%);
		clear(window.%\pc{$pos$}%);
	}
	
	%\textsc{Range}%<%\textsc{N}%> %\lstbtt{empty}%() {
		return %\textsc{Range}%(%\pc{$pos$}%, %\pc{$max$}% - %\pc{$pos$}%);
	}
	
	%\textsc{Boolean}% %\lstbtt{appendable}%(%\textsc{Sequence}%<%\textsc{N}%, %\textsc{V}%> sequence) {
		%\pc{If }%(%\pc{$pos$}% == %\pc{$max$}%) return false;
		%\pc{If }%(%\pc{$pos$}% < sequence.%\pc{$min$}%) return false;
		return (%\pc{$pos$}% < sequence.%\pc{$pos$}%);
	}

	void %\lstbtt{append}%(%\textsc{Sequence}%<%\textsc{N}%, %\textsc{V}%> sequence) {
		%\textsc{N}% from := max(sequence.%\pc{$min$}%, %\pc{$pos$}%);
		%\textsc{N}% to := min(sequence.%\pc{$pos$}%, %\pc{$max$}%);
		%\pc{For each \textsc{N} }%index%\pc{ in }%[from, to]: this[index] := sequence[index];
	}
	
	%\textsc{Sequence}%<%\textsc{N}%, %\textsc{V}%> %\lstbtt{sequence}%(%\textsc{Range}%<%\textsc{N}%> range) {
		/* Check range%\,%*/
		%\pc{If }%(range.%\pc{$from$}% < %\pc{$min$}%) return %\pc{nil}%;
		%\pc{If }%(range.%\pc{$from$}% >= %\pc{$pos$}%) return %\pc{nil}%;
		
		/* Determine output%\,%*/
		%\textsc{N}% to := min(range.%\pc{$from$}% + range.%\pc{$count$}%, %\pc{$pos$}%);
		%\textsc{Sequence}%<%\textsc{N}%, %\textsc{V}%> seq := %\textsc{Sequence}%(range.%\pc{$from$}%, to);
		%\pc{For each \textsc{N} }%index%\pc{ in }%[range.%\pc{$from$}%, to]{
			seq[index] := this[index];
		}
		return seq;
	}
}

typedef %\textsc{Windows}%<%\textsc{I}\pc{ is an }\textsc{ID}%, %\textsc{N}\pc{ is a }\textsc{Number}%, %\textsc{V}%>:%\\%%\textsc{Map}%<%\textsc{I}%, %\textsc{Window}%<%\textsc{N}%, %\textsc{V}%>>
class %\textsc{NumberOpinions}%<%\textsc{I}\pc{ is an }\textsc{ID}%, %\textsc{N}\pc{ is a }\textsc{Number}%>%\\%extends %\textsc{Map}%<%\textsc{I}%, %\textsc{N}%> {
	/* Constructor%\,%*/
	%\textsc{NumberOpinions}%() {
		%\pc{For each \textsc{I}} %id: this[id] := 0;
	}
	
	/* Operation%\,%*/
	%\textsc{N}% %\lstbtt{highest}%(%\textsc{Number}% threshold) {
		%\textsc{N}%[] ranking := %\pc{$values$ sorted in descending order}%;
		return ranking[threshold - 1];
	}
}

class %\textsc{ProgressOpinions}%<%\textsc{I}\pc{ is an }\textsc{ID}%,%\\%%\textsc{P}\pc{ is a }\textsc{Map}%<%\textsc{D}\pc{ is an }\textsc{ID}%, %\textsc{N}\pc{ is a }\textsc{Number}%>>%\\%extends %\textsc{Map}%<%\textsc{I}%, %\textsc{P}%> {
	/* Constructor%\,%*/
	%\textsc{ProgressOpinions}%() {
		%\pc{For each \textsc{I}} %key {
			this[key] := %\textsc{Map}%();
			%\pc{For each \textsc{D}} %id: this[key][id] := 0;
		}
	}
	
	/* Operation%\,%*/
	%\textsc{P}% %\lstbtt{highest}%(%\textsc{Number}% threshold) {
		%\textsc{P}% result := %\textsc{Map}%();
		%\pc{For each \textsc{D}} %id {
			%\textsc{NumberOpinions}%<%\textsc{I}%, %\textsc{N}%> opns := %\textsc{NumberOpinions}%();
			%\pc{For each \textsc{I}} %key: opns[key] := this[key][id];
			result[id] := opns.highest(threshold);
		}
		return result;
	}
}

class %\textsc{WindowOpinions}%<%\textsc{I}\pc{ is an }\textsc{ID}%,%\textsc{W}\pc{ is a }\textsc{Window}%<%\textsc{N}\pc{ is a }\textsc{Number}%, %\textsc{V}%>> extends %\textsc{Map}%<%\textsc{I}%, %\textsc{W}%> {
	/* Constructor%\,%*/
	%\textsc{WindowOpinions}%() {
		%\pc{For each \textsc{I}} %id: this[id] := %\textsc{Window}%();
	}
	
	/* Operations%\,%*/
	void %\lstbtt{fill}%(%\textsc{N}% to, %\textsc{V}% value) {
		%\pc{For each \textsc{I}} %id: this[id].fill(to, value);
	}

	void %\lstbtt{move}%(%\textsc{N}% min) {
		%\pc{For each \textsc{I}} %id: this[id].move(min);
	}
	
	void %\lstbtt{sync}%(%\textsc{Window}%<%\textsc{N}%, %\textsc{*}%> window) {
		%\pc{For each \textsc{I}} %id: this[id].sync(window);
	}
	
	void %\lstbtt{sync}%(%\textsc{N}% min) {
		%\pc{For each \textsc{I}} %id {
			this[id].move(min);
			this[id].clear(min);
		}
	}
	
	%\textsc{Number}% %\lstbtt{available}%(%\textsc{N}% index) {
		%\textsc{Number}% count := 0;
		%\pc{For each \textsc{I}} %id {
			%\pc{If }%(index < this[id].%\pc{$pos$}%) count++;
		}
		return count;
	}
	
	%\textsc{V}% %\lstbtt{any}%(%\textsc{N}% index) {
		%\pc{For each \textsc{I}} %id {
			%\pc{If }%(index < this[id].%\pc{$pos$}%) return this[id][index];
		}
		return %\pc{\possible}%;
}	}
class %\textsc{CommandID}% {
	/* State%\,%*/
	%\textsc{ClientID}% %\pc{$cid$}%;
	%\textsc{CommandNr}% %\pc{$xnr$}%;
	
	/* Constructor%\,%*/
	%\textsc{CommandID}%(%\textsc{ClientID}% c, %\textsc{CommandNr}% x) {
		%\pc{$cid$}% := c;
		%\pc{$xnr$}% := x;
	}
}

class %\textsc{Command}% {
	/* State%\,%*/
	%\textsc{CommandID}% %\pc{$xid$}%;
	%\textsc{Any}% %\pc{$op$}%;
	
	/* Auxiliary attributes%\,%*/
	%\textsc{ClientID}% %\pc{$cid$}% := %\pc{$xid$}%.%\pc{$cid$}%;
	%\textsc{CommandNr}% %\pc{$xnr$}% := %\pc{$xid$}%.%\pc{$xnr$}%;
	
	/* Constructor%\,%*/
	%\textsc{Command}%(%\textsc{ClientID}% cid, %\textsc{CommandNr}% xnr, %\textsc{Any}% op) {
		%\pc{$xid$}% := %\textsc{CommandID}%(cid, xnr);
		%\pc{$op$}% := op;
	}
}

class %\textsc{Result}% {
	/* State%\,%*/
	%\textsc{CommandID}% %\pc{$xid$}%;
	%\textsc{Any}% %\pc{$result$}%;
	
	/* Auxiliary attributes%\,%*/
	%\textsc{ClientID}% %\pc{$cid$}% := %\pc{$xid$}%.%\pc{$cid$}%;
	%\textsc{CommandNr}% %\pc{$xnr$}% := %\pc{$xid$}%.%\pc{$xnr$}%;
	
	/* Constructor%\,%*/
	%\textsc{Result}%(%\textsc{CommandID}% xid, %\textsc{Any}% result) {
		%\pc{$xid$}% := xid;
		%\pc{$result$}% := result;
	}
}

typedef %\textsc{CommandRange}%: %\textsc{Range}%<%\textsc{CommandNr}%>;
typedef %\textsc{CommandSequence}%<%\textsc{V}%>: %\textsc{Sequence}%<%\textsc{CommandNr}%, %\textsc{V}%>;
typedef %\textsc{CommandWindow}%<%\textsc{V}%>: %\textsc{Window}%<%\textsc{CommandNr}%, %\textsc{V}%>;
typedef %\textsc{CommandProgress}%: %\textsc{Map}%<%\textsc{ClientID}%, %\textsc{CommandNr}%>;
typedef %\textsc{CommandWindows}%<%\textsc{V}%>:%\\%%\textsc{Windows}%<%\textsc{ClientID}%, %\textsc{CommandNr}%, %\textsc{V}%>;%\vspace{2mm}\hrule\vspace{2mm}%
class %\textsc{Report}%<V> {
	/* State%\,%*/
	%\textsc{AgreementSequence}%<V> %\pc{$values$}%;
	%\textsc{View}% %\pc{$view$}%;
	
	/* Constructor%\,%*/
	%\textsc{Report}%(%\textsc{AgreementSequence}%<%\textsc{V}%> values, %\textsc{View}% view) {
		%\pc{$values$}% := values;
		%\pc{$view$}% := view;
	}
	
	/* Operations%\,%*/
	%\textsc{V}% %\lstbtt{get}%(%\textsc{AgreementNr}% a) {%\hfill%/* Accessed via %\lsttt{[]}% operator%\,%*/
		return %\pc{$values$}%[a];
	}
}

class %\textsc{AgreementWindow}%<%\textsc{V}%>%\\%extends %\textsc{Window}%<%\textsc{AgreementNr}%,%\,\,\textsc{V}%>%\,%{
	/* Operation%\,%*/
	%\textsc{Report}%<%\textsc{V}%> %\lstbtt{report}%(%\textsc{AgreementRange}% range, %\textsc{View}% view) {
		return %\textsc{Report}%(sequence(range), view);
	}
}

class %\textsc{Legacy}% {
	/* State%\,%*/
	%\textsc{View}% %\pc{$view$}%;
	%\textsc{Command}% %\pc{$command$}%;

	/* Constructor%\,%*/
	%\textsc{Legacy}%(%\textsc{View}% view, %\textsc{Command}% command) {
		%\pc{$view$}% := view;
		%\pc{$command$}% := command;
	}
}

typedef %\textsc{AgreementRange}%: %\textsc{Range}%<%\textsc{AgreementNr}%>;
typedef %\textsc{AgreementSequence}%<%\textsc{V}%>:%\\%%\textsc{Sequence}%<%\textsc{AgreementNr}%, %\textsc{V}%>;
typedef %\textsc{AgreementWindows}%<%\textsc{V}%>:%\\%%\textsc{Windows}%<%\textsc{CommitterID}%, %\textsc{AgreementNr}%, %\textsc{V}%>;%\vspace{2mm}\hrule\vspace{2mm}%
class %\textsc{Application}% {
	/* State%\,%*/
	%\textsc{State}% %\pc{$state$}%;
	
	/* Constructor%\,%*/
	%\textsc{Application}%() {
		%\pc{$state$}% := %\pc{initial state}%;
	}
	
	/* Operations%\,%*/
	%\textsc{Any}% %\lstbtt{execute}%(%\textsc{Any}% op) {
		%\pc{Apply }%op%\pc{ to $state$}%;
		return %\pc{result of operation }%op;
	}
}

class %\textsc{AgreementSnapshot}% {
	/* State%\,%*/
	%\textsc{AgreementNr}% %\pc{$anr$}%;
	%\textsc{CommandProgress}% %\pc{$complete$}%;

	/* Constructor%\,%*/
	%\textsc{AgreementSnapshot}%(%\textsc{AgreementNr}% a, %\textsc{CommandProgress}% p) {
		%\pc{$anr$}% := a;
		%\pc{$complete$}% := p;
	}
}

class %\textsc{ExecutionSnapshot}% {
	/* State%\,%*/
	%\textsc{AgreementSnapshot}% %\pc{$agreement$}%;
	%\textsc{State}% %\pc{$state$}%;
	%\textsc{Map}%<%\textsc{ClientID}%, %\textsc{Result}%> %\pc{$results$}%;
	
	/* Constructor%\,%*/
	%\textsc{ExecutionSnapshot}%(%\textsc{AgreementNr}\,%a, %\textsc{CommandProgress}\,%p, %\textsc{State}% s, %\textsc{Map}%<%\textsc{ClientID}%, %\textsc{Result}%> r) {
		%\pc{$agreement$}% := %\textsc{AgreementSnapshot}%(a, p);
		%\pc{$state$}% := s;
		%\pc{$results$}% := r;
	}
}

typedef %\textsc{CheckpointNr}%: %\textsc{Number}%;

%\textsc{CheckpointNr}% anr2cnr(%\textsc{AgreementNr}% anr) {
	return %\pc{$\lfloor$}%a / CHECKPOINT_INTERVAL%\pc{$\rfloor$}%;
}

%\textsc{AgreementNr}% cnr2anr(%\textsc{CheckpointNr}% cnr) {
	return cnr * CHECKPOINT_INTERVAL;
}

typedef %\textsc{CheckpointRange}%: %\textsc{Range}%<%\textsc{CheckpointNr}%>;
typedef %\textsc{CheckpointSequence}%<%\textsc{V}%>:%\\%%\textsc{Sequence}%<%\textsc{CheckpointNr}%, %\textsc{V}%>;
typedef %\textsc{CheckpointWindow}%<%\textsc{V}%>: %\textsc{Window}%<%\textsc{CheckpointNr}%, %\textsc{V}%>;

class %\textsc{Checkpoint}%<%\textsc{S}%> {
	/* State%\,%*/
	%\textsc{CheckpointNr}% %\pc{$cnr$}%;
	%\textsc{S}% %\pc{$snapshot$}%;
	
	/* Auxiliary attributes%\,%*/
	... := %\pc{attributes of $snapshot$}%;
	
	/* Constructor%\,%*/
	%\textsc{Checkpoint}%(%\textsc{CheckpointNr}% c, %\textsc{S}% snapshot) {
		%\pc{$cnr$}% := c;
		%\pc{$snapshot$}% := snapshot;
	}
}

typedef %\textsc{AgreementCheckpoint}%:%\\%%\textsc{Checkpoint}%<%\textsc{AgreementSnapshot}%>;
typedef %\textsc{ExecutionCheckpoint}%:%\\%%\textsc{Checkpoint}%<%\textsc{ExecutionSnapshot}%>;

\end{lstlisting}

\subsection{Client}

\begin{lstlisting}[aboveskip=1mm]
class %\textsc{Client}% {
	/* State%\,%*/
	%\textsc{CommandWindow}%<%\textsc{Commmand}%> %\pc{$commands$}%;
	
	/* Initialization%\,%*/
	%\pc{On system start}% {
		%\pc{$commands$}% := %\textsc{CommandWindow}%();
	}

	/* Services%\,%*/
	%\textsc{CommandSequence}%<%\textsc{Command}%> %\lstbtt{getCommands}%(%\textsc{CommandRange}% r) {
		return %\pc{$commands$}%.sequence(r);
	}
	
	/* Command submission%\,%*/
	%\textsc{Boolean}% %\lstbtt{invoke}%(%\textsc{Any}% a) {
		/* Check state%\,%*/
		%\pc{If }%(%\pc{$commands$}%.%\pc{$pos$}% == %\pc{$commands$}%.%\pc{$max$}%) return false;
		
		/* Update state%\,%*/
		%\pc{$commands$}%[%\pc{$commands$}%.%\pc{$pos$}%] := %\textsc{Command}%(this.%\pc{$id$}%,%\\%%\pc{$commands$}%.%\pc{$pos$}%, a);
		return true;
	}
	
	/* Periodic tasks%\,%*/
	%\pc{Periodically}% {
		/* Check state%\,%*/
		%\textsc{CommandNr}% count := %\pc{$commands$}%.%\pc{$pos$}% - %\pc{$commands$}%.%\pc{$min$}%;
		%\pc{If }%(count == 0) return;
		
		/* Fetch results%\,%*/
		%\textsc{CommandRange}% r := %\textsc{CommandRange}%(%\pc{$commands$}%.%\pc{$min$}%,count);
		%\pc{For each \textsc{Executor} }%exr: exr.getResults(this.%\pc{$id$}%, r);
	}
	
	%\pc{On receiving \textsc{CommandSequence}}%<%\textsc{Result}%> rs%\pc{ from \textsc{Executor} }%exr {
		/* Check input%\,%*/
		%\pc{If }%(rs.%\pc{$pos$}% <= %\pc{$commands$}%.%\pc{$min$}%) return;
		%\pc{If }%(rs %\pc{is not authentic}%) return;
		
		/* Deliver results%\,%*/
		%\pc{For each \textsc{CommandNr} }%x%\pc{ in }%[%\pc{$commands$}%.%\pc{$min$}%, rs.%\pc{$pos$}%]:%\\%%\pc{Deliver }%rs[x]%\pc{ to user}%;
		
		/* Move window%\,%*/
		%\pc{$commands$}%.move(rs.%\pc{$pos$}%);
	}
}
\end{lstlisting}

\subsection{Front End}
\label{spec:fre}

\begin{lstlisting}[aboveskip=1mm]
class %\textsc{FrontEnd}% extends %\textsc{CompletionObserver}% {
	/* State%\,%*/
	%\textsc{CommandWindows}%<%\textsc{Command}%> %\pc{$commands$}%;
	%\textsc{CommandProgress}% %\pc{$submitted$}%;
	
	/* Initialization%\,%*/
	%\pc{On system start}% {
		%\pc{$commands$}% := %\textsc{CommandWindows}%();
		%\pc{$submitted$}% := %\textsc{CommandProgress}%();
	}
	
	/* Window synchronization%\,%*/
	%\pc{On} \textsc{CommandProgress} \pc{$complete$ change}% {
		%\pc{For each \textsc{ClientID} }%clt {
			%\pc{$commands$}%[clt].move(%\pc{complete}%[clt]);
			%\pc{submitted}%[clt] := max(%\pc{submitted}%[clt], %\pc{complete}%[clt]);
		}
	}
	
	/* Services%\,%*/
	%\textsc{CommandSequences}%<%\textsc{Command}%> %\lstbtt{getCommands}%(%\textsc{CommandRanges}% rs) {
		return %\pc{$commands$}%.sequences(rs);
	}

	%\textsc{CommandProgress}% %\lstbtt{getSubmitted}%() {
		return %\pc{$submitted$}%;
	}
	
	/* Periodic tasks%\,%*/
	%\pc{Periodically}% {
		/* Fetch client commands%\,%*/
		%\pc{For each \textsc{Client} }%clt {
			%\textsc{CommandRange}% r := %\pc{$commands$}%[clt].empty();
			%\pc{If }%(r.%\pc{$count$}% > 0) clt.getCommands(r);
		}
		
		/* Fetch front-end commands%\,%*/
		%\pc{For each \textsc{FrontEnd} }%fre {%\label{spec:fre:fetch-start}%
			%\textsc{CommandRanges}% rs := %\pc{$commands$}%.empty();
			%\pc{If }%(rs.size() > 0) fre.getCommands(rs);
		}%\label{spec:fre:fetch-end}%
	}
	
	%\pc{On receiving \textsc{CommandSequence}}%<%\textsc{Command}%>%\,%xs%\pc{ from \textsc{Client} }%clt {
		store(xs, clt);
	}

	%\pc{On receiving \textsc{CommandSequences}}%<%\textsc{Command}%>%\,%xss%\pc{ from \textsc{FrontEnd} }%fre {
		%\pc{For each }%%\textsc{ClientID}% clt: store(xss[clt], clt);
	}

	/* Auxiliary method%\,%*/
	void store(%\textsc{CommandSequence}%<%\textsc{Command}%> xs, %\textsc{ClientID}% clt) {
		/* Check input%\,%*/
		%\pc{If }%(%\pc{$commands$}%[clt].appendable(xs) == false) return;

		/* Check input values%\,%*/
		%\textsc{CommandNr}% from := max(xs.%\pc{$min$}%, %\pc{$commands$}%[clt].%\pc{$pos$}%);
		%\textsc{CommandNr}% to := min(xs.%\pc{$pos$}%, %\pc{$commands$}%[clt].%\pc{$max$}%);
		%\pc{For each }%%\textsc{CommandNr}% x%\pc{ in }%[from, to] {
			%\pc{If }%(xs[x].validate(clt, x) == false) return;	
		}

		/* Check input authenticity%\,%*/
		%\pc{If }%(xs %\pc{is not authentic}%) return;

		/* Update output%\,%*/
		%\pc{$commands$}%[clt].append(xs);
		%\pc{$submitted$}%[clt] := max(%\pc{$submitted$}%[clt], %\pc{$commands$}%[clt].%\pc{$pos$}%);
	}
}
\end{lstlisting}

\subsection{Proposer}

\begin{lstlisting}[aboveskip=1mm]
class %\textsc{Proposer}% extends %\textsc{AgreementObserver}%, %\textsc{CompletionObserver}%, %\textsc{ViewObserver}% {
	/* Main State%\,%*/
	%\textsc{AgreementWindow}%<%\textsc{Command}%> %\pc{$proposals$}%;
	%\textsc{CommandProgress}% %\pc{$proposed$}%;
	%\textsc{Mode}% %\pc{$mode$}%;%\hfill%/* %\lsttt{NORMAL}%, %\lsttt{VIEW\_CHANGE}%, or %\lsttt{IDLE}\,%*/


	/* Initialization%\,%*/
	%\pc{On system start}% {
		%\pc{$proposals$}% := %\textsc{AgreementWindow}%();
		%\pc{$proposed$}% := %\textsc{CommandProgress}%();
		%\pc{$mode$}% := ite(this == %\pc{$view$}%.%\pc{$proposer$}%, NORMAL, IDLE);
	}
	
	/* Window synchronization%\,%*/
	%\pc{On} \textsc{AgreementNr} \pc{$agreed$ change}% {
		/* Move windows%\,%*/
		%\pc{$proposals$}%.move(%\pc{$agreed$}%);
		shift();%\hfill%/* Mode-specific implementations%\,%*/
	}

	%\pc{On} \textsc{CommandProgress} \pc{$complete$ change}% {
		completed();%\hfill%/* Mode-specific implementations%\,%*/
	}
	
	/* View change%\,%*/
	%\pc{On \textsc{View}}% %\pc{$view$ change}% {
		/* Reset output%\,%*/
		%\pc{$proposals$}%.reset();
		
		/* Switch mode%\,%*/
		%\pc{If }%(this != %\pc{$view$}%.%\pc{$proposer$}%) {
			%\pc{$mode$}% := IDLE;
		} %\pc{else}% {
			%\pc{$proposed$}% := %\pc{$complete$}%;
			%\pc{$mode$}% := VIEW_CHANGE;
		}
	}
	
	/* Services%\,%*/
	%\textsc{Report}%<%\textsc{Command}%> %\lstbtt{getProposals}%(%\textsc{AgreementRange}% r,%\textsc{View}% v) {
		%\pc{If }%((v == %\pc{$view$}%) %\pc{$\wedge$}% (this == v.%\pc{$proposer$}%)) {
			return %\pc{$proposals$}%.report(r, v);
		}
	}%\vspace{2mm}\hrule\pc{\vspace{2mm}\begin{center}{\lsttt{NORMAL}}\end{center}}\hrule\vspace*{2mm}%
	/* Mode-specific normal-case state%\,%*/
	%\textsc{CommandWindows}%<%\textsc{Command}%> %\pc{$commands$}%;
	
	/* Initialization%\,%*/
	%\pc{On system start}% {
		%\pc{$commands$}% := %\textsc{CommandWindows}%();
	}

	%\pc{On mode start}% {
		/* Move commands%\,%*/
		%\pc{For each \textsc{ClientID}}% clt {
			%\pc{$commands$}%[clt].move(%\pc{$complete$}%[clt]);
		}
		update();
	}
	
	/* Window synchronization%\,%*/
	void shift() {
		update();
	}

	void completed() {
		/* Move commands%\,%*/
		%\pc{For each \textsc{ClientID}}% clt {
			%\pc{$commands$}%[clt].move(%\pc{$complete$}%[clt]);
		}
		update();
	}
	
	/* Periodic command%\,%*/
	%\pc{Periodically}% {
		%\pc{For each \textsc{FrontEnd} }%fre {
			%\textsc{CommandRanges}% rs := %\pc{$commands$}%.empty();
			%\pc{If }%(rs.size() > 0) fre.getCommands(rs);
		}
	}
	
	%\pc{On receiving \textsc{CommandSequence}}%<%\textsc{Command}%>%\,%xss%\pc{ from \textsc{FrontEnd} }%fre {
		/* Store input%\,%*/
		%\pc{For each }%%\textsc{ClientID}% clt {
			/* Check input%\,%*/
			%\pc{If }%(%\pc{$commands$}%[clt].appendable(xss[clt]) == false) continue;
			
			/* Check input values%\,%*/
			%\textsc{CommandNr}% from := max(xss[clt].%\pc{$min$}%, %\pc{$commands$}%[clt].%\pc{$pos$}%);
			%\textsc{CommandNr}% to := min(xss[clt].%\pc{$pos$}%, %\pc{$commands$}%[clt].%\pc{$max$}%);
			%\pc{For each }%%\textsc{CommandNr}% x%\pc{ in }%[from, to] {
				%\pc{If }%(xss[clt][x].validate(clt, x) == false) return;
			}

			/* Check input authenticity%\,%*/
			%\pc{If }%(xss %\pc{is not authentic}%) return;

			/* Store input%\,%*/
			%\pc{$commands$}%[clt].append(xss[clt]);
		}
		
		/* Update output%\,%*/
		update();
	}
	
	/* Auxiliary method%\,%*/
	void update() {
		%\pc{For each \textsc{AgreementNr} }%a%\pc{ in }%[%\pc{$proposals$}%.%\pc{$pos$}%, %\pc{$proposals$}%.%\pc{$max$}%] {
			/* Determine output value%\,%*/
			%\textsc{Command}% x := %\pc{nil}%;
			%\pc{For each }%%\textsc{ClientID}% clt%\pc{ in random order}% {
				x := %\pc{$commands$}%[clt][%\pc{$proposed$}%[clt]];
				%\pc{If }%(x != %\pc{nil}%) break;
			}
			%\pc{If }%(x == %\pc{nil}%) break;

			/* Update output value%\,%*/
			%\pc{$proposals$}%[a] := x
			%\pc{$proposed$}%[x.%\pc{$cid$}%] := x.%\pc{$xnr$}% + 1
		}
	}%\vspace{2mm}\hrule\pc{\begin{center}{\lsttt{VIEW\_CHANGE}}\end{center}}\hrule\vspace*{2mm}%
	/* Mode-specific view-change state%\,%*/
	%\textsc{WindowOpinions}%<%\textsc{CommitterID}%, %\textsc{AgreementWindow}%<%\textsc{Legacy}%>> %\pc{$legacies$}%;
	
	/* Initialization%\,%*/
	%\pc{On system start}% {
		%\pc{$legacies$}% := %\textsc{WindowOpinions}%();
	}

	%\pc{On mode start or restart}% {
		%\pc{$legacies$}%.sync(%\pc{$proposals$}%);
	}
	
	/* Window synchronization%\,%*/
	void shift() {
		%\pc{$legacies$}%.move(%\pc{$agreed$}%);
	}
	%\newpage%
	void completed() {
		/* Do nothing%\,%*/
	}

	/* Periodic tasks%\,%*/
	%\pc{Periodically}% {
		%\pc{For each \textsc{Committer} }%cmr {
			%\textsc{AgreementRange}% r := %\pc{$legacies$}%[cmr.%\pc{$id$}%].empty();
			%\pc{If }%(r.%\pc{$count$}% > 0) cmr.getLegacies(r, %\pc{$view$}%);
		}
	}
	
	%\pc{On receiving \textsc{Report}}%<%\textsc{Legacy}%>%\,%rs%\pc{ from \textsc{Committer} }%cmr {
		/* Check input%\,%*/
		%\pc{If }%(rs.%\pc{$view$}% != %\pc{$view$}%) return;
		%\pc{If }%(%\pc{$legacies$}%[cmr].appendable(rs.%\pc{$values$}%) == false) return;
		%\pc{If }%(rs %\pc{is not authentic}%) return;
		
		/* Store input%\,%*/
		%\pc{$legacies$}%[cmr].append(rs.%\pc{$values$}%);
		
		/* Update output%\,%*/
		%\pc{For each \textsc{AgreementNr} }%a%\pc{ in }%[%\pc{$proposals$}%.%\pc{$pos$}%, %\pc{$proposals$}%.%\pc{$max$}%] {
			/* Check state%\,%*/
			%\pc{If }%(%\pc{$legacies$}%.available(a) < F+1) break;
			
			/* Determine output value%\,%*/
			%\textsc{Legacy}% l := %\textsc{Legacy}%(-1, %\pc{\emptyslot}%);
			%\pc{For each \textsc{CommitterID} }%i {
				%\pc{If }%(%\pc{$legacies$}%[i][a] == %\pc{nil}%) continue;
				%\pc{If }%(%\pc{$legacies$}%[i][a].%\pc{$command$}% == %\pc{\emptyslot}%) continue;
				%\pc{If }%(%\pc{$legacies$}%[i][a].%\pc{$view$}% > l.%\pc{$view$}%) l := %\pc{$legacies$}%[i][a];
			}
			
			/* Switch mode if the output is complete%\,%*/
			%\pc{If }%(l.%\pc{$command$}% == %\pc{\emptyslot}%) {
				mode := NORMAL;
				return;
			}
			
			/* Update output value%\,%*/
			%\textsc{Command}% x := l.%\pc{$command$}%;
			%\pc{$proposals$}%[a] := x;
			%\pc{$proposed$}%[x.%\pc{$cid$}%] := x.%\pc{$xnr$}% + 1;
		}
		
		/* Fill unused input slots%\,%*/
		%\pc{$legacies$}%.fill(%\pc{$proposals$}%.%\pc{$pos$}%, %\pc{\emptyslot}%);
	}%\vspace{2mm}\hrule\pc{\begin{center}{\lsttt{IDLE}}\end{center}}\hrule\vspace*{2mm}%
	/* Window synchronization%\,%*/
	void shift() {
		/* Do nothing%\,%*/
	}

	void completed() {
		/* Do nothing%\,%*/
	}
}
\end{lstlisting}

\subsection{Committer}

\begin{lstlisting}[aboveskip=1mm]
class %\textsc{Committer}% extends %\textsc{AgreementObserver}%, %\textsc{ViewObserver}% {
	/* State%\,%*/
	%\textsc{AgreementWindow}%<%\textsc{Command}%> %\pc{$commits$}%;
	%\textsc{AgreementWindow}%<%\textsc{Legacy}%> %\pc{$legacies$}%;
	
	/* Initialization%\,%*/
	%\pc{On system start}% {
		%\pc{$commits$}% := %\textsc{AgreementWindow}%();
		%\pc{$legacies$}% := %\textsc{AgreementWindow}%();
		%\pc{$legacies$}%.fill(%\pc{$legacies$}%.%\pc{$max$}%, %\textsc{Legacy}%(-1, %\pc{\emptyslot}%));
	}
	
	/* Window synchronization%\,%*/
	%\pc{On} \textsc{AgreementNr} \pc{$agreed$ change}% {
		%\pc{$commits$}%.move(%\pc{$agreed$}%);
		%\pc{$legacies$}%.move(%\pc{$agreed$}%, %\textsc{Legacy}%(%\pc{$view$}% - 1, %\pc{\emptyslot}%));
	}
	
	/* View change%\,%*/
	%\pc{On} \textsc{View} \pc{$view$ change from $old\_view$}% {
		/* Update state%\,%*/
		%\pc{For each \textsc{AgreementNr} }%a%\pc{ in }%[%\pc{$commits$}%.%\pc{$min$}%, %\pc{$commits$}%.%\pc{$pos$}%] {
			%\pc{$legacies$}%[a] := %\textsc{Legacy}%(%\pc{$old\_view$}%, %\pc{$commits$}%[a]);
		}
		
		/* Reset output%\,%*/
		%\pc{$commits$}%.reset();
	}
	
	/* Services%\,%*/
	%\textsc{Report}%<%\textsc{Command}%> %\lstbtt{getCommits}%(%\textsc{AgreementRange}% r, %\textsc{View}% v) {
		%\pc{If }%(v == %\pc{$view$}%) return %\pc{$commits$}%.report(r, v);
	}
	
	%\textsc{Report}%<%\textsc{Legacy}%> %\lstbtt{getLegacies}%(%\textsc{AgreementRange}% r, %\textsc{View}% v) {
		%\pc{If }%(v == %\pc{$view$}%) return %\pc{$legacies$}%.report(r, v);
	}
	
	/* Periodic tasks%\,%*/
	%\pc{Periodically}% {
		%\textsc{AgreementRange}% r := %\pc{$commits$}%.empty();
		%\pc{If }%(r.%\pc{$count$}% > 0) %\pc{$view$}%.%\pc{$proposer$}%.getProposals(r, %\pc{$view$}%);
	}
	
	%\pc{On receiving }\textsc{Report}%<%\textsc{Command}%>%\,%xs%\pc{ from \textsc{Proposer} }%pps {
		/* Check input%\,%*/
		%\pc{If }%(xs.%\pc{$view$}% != %\pc{$view$}%) return;
		%\pc{If }%(%\pc{$commits$}%.appendable(xs.%\pc{$values$}%) == false) return;
		%\pc{If }%(xs %\pc{is not authentic}%) return;
		
		/* Update output%\,%*/
		%\pc{$commits$}%.append(xs.%\pc{$values$}%);
	}
}
\end{lstlisting}

\subsection{Executor}

\begin{lstlisting}[aboveskip=1mm]
class %\textsc{Executor}% extends %\textsc{AgreementObserver}%, %\textsc{ViewObserver}% {
	/* State%\,%*/
	%\textsc{Application}% %\pc{$application$}%;
	%\textsc{AgreementNr}% %\pc{$next$}%;
	%\textsc{CommandWindows}%<%\textsc{Result}%> %\pc{$results$}%;
	%\textsc{CommandProgress}% %\pc{$complete$}%;

	/* Sync State%\,%*/
	%\textsc{CheckpointWindow}%<%\textsc{ExecutionSnapshot}%> %\pc{$snapshots$}%;
	%\textsc{Mode}% %\pc{$mode$}%;

	/* Initialization%\,%*/
	%\pc{On system start}% {
		%\pc{$application$}% := %\textsc{Application}%(); 
		%\pc{$next$}% := 0;
		%\pc{$results$}% := %\textsc{CommandWindows}%();
		%\pc{$complete$}% := %\textsc{CommandProgress}%();
		%\pc{$snapshots$}% := %\textsc{CheckpointWindow}%()%\\%%\pc{with $snapshots$}%[0] = %\textsc{ExecutionSnapshot}%();
		%\pc{$mode$}% := NORMAL;
	}

	/* Window synchronization%\,%*/
	%\pc{On} \textsc{AgreementNr} \pc{$agreed$ change}% {
		/* Switch mode if necessary%\,%*/
		%\pc{If}% (%\pc{$next$}% < %\pc{$agreed$}%) {
			%\pc{$mode$}% := SYNC;
			return;
		}

		/* Move window%\,%*/
		%\pc{$snapshots$}%.move(anr2cnr(%\pc{$agreed$}%));
		shift();%\hfill%/* Mode-specific implementations%\,%*/
	}

	/* View change%\,%*/
	%\pc{On} \textsc{View} \pc{$view$ change}% {
		viewChange();%\hfill%/* Mode-specific implementations%\,%*/
	}

	/* Services%\,%*/
	%\textsc{CommandSequence}%<%\textsc{Result}%> getResults(%\textsc{ClientID}% clt, %\textsc{CommandRange}% r) {
		return %\pc{$results$}%[clt].sequence(r);
	}

	%\textsc{ExecutionCheckpoint}% getExecutionCheckpoint(%\textsc{CheckpointNr}% c){
		return %\textsc{ExecutionCheckpoint}%(c, %\pc{$snapshots$}%[c]);
	}

	%\textsc{AgreementNr}% getAgreed() {
		return %\pc{$snapshots$}%[%\pc{$snapshots.pos$}% - 1].%\pc{$anr$}%;
	}

	%\textsc{CommandProgress}% getComplete() {
		return %\pc{$snapshots$}%[%\pc{$snapshots.pos$}% - 1].%\pc{$complete$}%;
	}

	%\textsc{CommandProgress}% getProcessed() {
		return %\pc{$complete$}%;
	}%\vspace{2mm}\hrule\vspace{0.75em}\pc{\begin{center}{\lsttt{NORMAL}}\end{center}}\hrule\vspace*{2mm}%
	/* Mode-specific state%\,%*/
	%\textsc{WindowOpinions}%<%\textsc{CommitterID}%,%\\%%\textsc{AgreementWindow}%<%\textsc{Command}%>> %\pc{$commits$}%;

	/* Initialization%\,%*/
	%\pc{On system start}% {
		%\pc{$commits$}% := %\textsc{WindowOpinions}%();
	}

	%\pc{On mode start or restart}% {
		%\pc{$commits$}%.move(%\pc{$next$}%);
	}

	/* Window synchronization%\,%*/
	void shift() {
		%\pc{$commits$}%.move(%\pc{$agreed$}%);
	}

	/* View change%\,%*/
	void viewChange() {
		%\pc{For each \textsc{CommitterID}}% cmr: %\pc{$commits$}%[cmr].clear(%\pc{$next$}%);
	}

	/* Periodic tasks%\,%*/
	%\pc{Periodically}% {
		%\pc{For each} \textsc{Committer}% cmr {
			%\textsc{AgreementRange}% r := %\pc{commits}%[cmr.%\pc{$id$}%].empty();
			%\pc{If}% (r.%\pc{$count$}% > 0) cmr.getCommits(r, %\pc{$view$}%);
		}
	}

	/* Inputs%\,%*/
	%\pc{On receiving} \textsc{Report}%<%\textsc{Command}%> xs %\pc{from} \textsc{Committer}% cmr {
		/* Check input%\,%*/
		%\pc{If}% (xs.%\pc{$view$}% != %\pc{$view$}%) return;
		%\pc{If}% (%\pc{$commits$}%[cmr].appendable(xs.%\pc{$values$}%) == false) return;
		%\pc{If}% (xs %\pc{is not authentic}%) return;

		/* Store input%\,%*/
		%\pc{$commits$}%[cmr].append(xs.%\pc{$values$}%);

		/* Update state%\,%*/
		%\pc{For each} \textsc{AgreementNr}% a %\pc{in}% [%\pc{$next$}%, %\pc{$commits$}%[cmr].%\pc{$pos$}%] {
			/* Check state%\,%*/
			%\pc{If}% (%\pc{$commits$}%.available(a) < F+1) break;

			/* Determine state value%\,%*/
			%\textsc{Command}% x := %\pc{$commits$}%.any(a);
			/* Update state and output if necessary%\,%*/
			%\pc{If}% (x.%\pc{$xnr$}% >= %\pc{$complete$}%.[x.%\pc{$cid$}%]) {
				%\textsc{Any}% result := %\pc{$application$}%.execute(x.%\pc{$op$}%);
				%\pc{$results$}%[x.%\pc{$cid$}%].move(x.%\pc{$xnr$}% - %\pc{$results$}%[x.%\pc{$cid$}%].%\pc{$capacity$}% + 1);
				%\pc{$results$}%[x.%\pc{$cid$}%] := %\textsc{Result}%(x.%\pc{$xid$}%, result);
				%\pc{$complete$}%[x.%\pc{$cid$}%] := x.%\pc{$xnr$}% + 1;
			}
			%\pc{$next$}% := a + 1;

			/* Create snapshot if necessary%\,%*/
			%\textsc{CheckpointNr}% c := anr2cnr(%\pc{$next$}%);
			%\pc{If}% (c == %\pc{$snapshots.pos$}%) {
				%\pc{$snapshots$}%[c] := %\textsc{ExecutionSnapshot}%(%\pc{$next$}%, %\pc{$complete$}%, %\pc{$application.state$}%, %\pc{$results$}%);
			}
		}

		/* Fill unused input slots%\,%*/
		%\pc{$commits$}%.fill(%\pc{$next$}%, %\pc{\emptyslot}%);
	}%\vspace{2mm}\hrule\pc{\begin{center}{\lsttt{SYNC}}\end{center}}\hrule\vspace*{2mm}%
	/* Window synchronization%\,%*/
	void shift() {
		/* Do nothing%\,%*/
	}

	/* View change%\,%*/
	void viewChange() {
		/* Do nothing%\,%*/
	}

	/* Periodic tasks%\,%*/
	%\pc{Periodically}% {
		%\textsc{CheckpointNr}% c := anr2cnr(%\pc{$agreed$}%);
		%\pc{For each \textsc{Executor}}% exr {
			exr.getExecutionCheckpoint(c);
		}
	}

	/* Inputs%\,%*/
	%\pc{On receiving \textsc{ExecutionCheckpoint}}% c %\pc{from \textsc{Executor}}% exr {
		/* Check input%\,%*/
		%\pc{If}% (c.%\pc{$anr$}% < %\pc{$agreed$}%) return;
		%\pc{If}% (c %\pc{is not authentic}%) return;

		/* Store input and switch mode%\,%*/
		%\pc{$application.state$}% := c.%\pc{$state$}%;
		%\pc{$next$}% := c.%\pc{$anr$}%;
		%\pc{$results$}% := c.%\pc{$results$}%;
		%\pc{$complete$}% := c.%\pc{$complete$}%;
		%\pc{$snapshots$}%.move(c.%\pc{$cnr$}%);
		%\pc{$snapshots$}%[c.%\pc{$cnr$}%] := c.%\pc{$snapshot$}%;
		%\pc{$mode$}% := NORMAL;
	}
}
\end{lstlisting}

\vspace{-1.5mm}

\subsection{Controller}
\label{spec:ctr}

\vspace{-.5mm}

\begin{lstlisting}[aboveskip=1mm]
class %\textsc{Controller}% extends %\textsc{ViewObserver}% {
	/* State%\,%*/
	%\textsc{Timestamp}% %\pc{$deadline$}%;
	%\textsc{Mode}% %\pc{$mode$}%;%\hfill%/* %\lsttt{NORMAL}% or %\lsttt{IDLE}\,%*/
	
	/* Initialization%\,%*/
	%\pc{On system start}% {
		%\pc{$deadline$}% := %\pc{$\infty$}%;
		%\pc{$mode$}% := NORMAL;
	}
		
	/* View change%\,%*/
	%\pc{On} \textsc{View} \pc{$view$ change}% {
		%\pc{$mode$}% := NORMAL;
	}
	
	/* Services%\,%*/
	%\textsc{View}% %\lstbtt{getView}%() {
		return ite(%\pc{$mode$}% == NORMAL, %\pc{$view$}%, %\pc{$view$}% + 1);
	}%\vspace{2mm}\hrule\pc{\vspace{2mm}\begin{center}{\lsttt{NORMAL}}\end{center}}\hrule\vspace*{2mm}%
	/* Mode-specific input state%\,%*/
	%\textsc{ProgressOpinions}%<%\textsc{ExecutorID}%, %\textsc{CommandProgress}%> %\pc{$submitted$}%;
	%\textsc{CommandProgress}% %\pc{$target$}%;
	%\textsc{ProgressOpinions}%<%\textsc{FrontEndID}%, %\textsc{CommandProgress}%> %\pc{$processed$}%;
	%\textsc{CommandProgress}% %\pc{$actual$}%;
	
	/* Mode-specific control state%\,%*/
	%\textsc{Map}%<%\textsc{ClientID}%, %\textsc{Timestamp}%> %\pc{$timestamps$}%;
	%\textsc{Timeout}% %\pc{$timeout$}%;
	
	/* Initialization%\,%*/
	%\pc{On system start}% {
		%\pc{$submited$}% := %\textsc{ProgressOpinions}%();
		%\pc{$target$}% := %\textsc{CommandProgress}%();
		%\pc{$processed$}% := %\textsc{ProgressOpinions}%();
		%\pc{$actual$}% := %\textsc{CommandProgress}%();
		%\pc{For each \textsc{ClientID} }%clt: %\pc{$timestamps$}%[clt] := 0;
		%\pc{$timeout$}% := CONTROLLER_TIMEOUT;
	}
	
	%\pc{On mode start or restart}% {
		%\pc{For each \textsc{ClientID} }%clt: %\pc{$timestamps$}%[clt] := %\pc{now}%;
		deadline();
	}
	
	/* Periodic tasks%\,%*/
	%\pc{Periodically}% {
		/* Check timeout expiration%\,%*/
		%\pc{If }%(%\pc{$deadline$}% <= %\pc{now}%) {
			%\pc{$timeout$}% := %\pc{$timeout$}% * 2;%\label{spec:ctr:back-off}%
			%\pc{$mode$}% := IDLE;
			return;
		}
		
		/* Fetch submission progresses%\,%*/
		%\pc{For each \textsc{FrontEnd} }%fre: fre.getSubmitted();
		
		/* Fetch finalization progresses%\,%*/
		%\pc{For each \textsc{Executor} }%exr: exr.getProcessed();
	}

	%\pc{On receiving \textsc{CommandProgress} }%ps%\pc{ from \textsc{FrontEnd} }%fre {
		/* Check input%\,%*/
		%\pc{If }%(ps %\pc{$\preceq$}% %\pc{$submitted$}%[fre]) return;
		%\pc{If }%(ps %\pc{is not authentic}%) return;
		
		/* Store input%\,%*/
		%\pc{$submitted$}%[fre] := ps;
		
		/* Update state%\,%*/
		%\textsc{CommandProgress}% vs := %\pc{$submitted$}%.highest(F+1);%\label{spec:ctr:target-start}%
		%\pc{For each \textsc{ClientID} }%clt {
			%\pc{If }%(vs[clt] <= %\pc{$target$}%[clt]) continue;
			%\pc{$target$}%[clt] := vs[clt];
			%\pc{$timestamps$}%[clt] := %\pc{now}%;
		}%\label{spec:ctr:target-end}%
		deadline();
	}

	%\pc{On receiving \textsc{CommandProgress} }%ps%\pc{ from \textsc{Executor} }%exr {
		/* Check input%\,%*/
		%\pc{If }%(ps %\pc{$\preceq$}% %\pc{$processed$}%[exr]) return;
		%\pc{If }%(ps %\pc{is not authentic}%) return;
		
		/* Store input%\,%*/
		%\pc{$processed$}%[exr] := ps;
		
		/* Update state%\,%*/
		%\textsc{CommandProgress}% vs := %\pc{$processed$}%.highest(F+1);
		%\pc{If }%(%\pc{$actual$ $\prec$}% vs) %\pc{$timeout$}% := CONTROLLER_TIMEOUT;
		%\pc{$actual$}% := vs;
		deadline();
	}
	
	/* Auxiliary method%\,%*/
	void deadline() {
		/* Determine time%\,%*/
		%\pc{$deadline$}% := %\pc{$\infty$}%;
		%\pc{For each \textsc{ClientID} }%clt {
			%\pc{If }%(%\pc{$target$}%[clt] <= %\pc{$actual$}%[clt]) continue;
			%\pc{$deadline$}% := min(%\pc{$deadline$}%, %\pc{$timestamps$}%[clt] + %\pc{$timeout$}%);
		}
	}%\vspace{2mm}\hrule\pc{\begin{center}{\lsttt{IDLE}}\end{center}}\hrule\vspace*{2mm}%
	/* Do nothing%\,%*/
}
\end{lstlisting}

\vspace{-1.5mm}

\subsection{Agreement Monitor and Observer}

\vspace{-.5mm}

\begin{lstlisting}[aboveskip=1mm]
class %\textsc{AgreementMonitor}% {
	/* State%\,%*/
	%\textsc{NumberOpinions}%<%\textsc{ExecutorID}%, %\textsc{AgreementNr}%> %\pc{$executors$}%;
	%\textsc{AgreementNr}% %\pc{$threshold$}%;
	
	/* Initialization%\,%*/
	%\pc{On system start}% {
		%\pc{$executors$}% :=  %\textsc{NumberOpinions}%();
		%\pc{$threshold$}% := 0;
	}
	
	/* Services%\,%*/
	%\textsc{AgreementNr}% %\lstbtt{getThreshold}%() {
		return %\pc{$threshold$}%;
	}
	
	/* Periodic tasks%\,%*/
	%\pc{Periodically}% {
		/* Fetch executor thresholds%\,%*/
		%\pc{For each \textsc{Executor} }%exr: exr.getAgreed();
		
		/* Fetch monitor thresholds%\,%*/
		%\pc{For each \textsc{AgreementMonitor} }%agm: agm.getThreshold();
	}
	
	%\pc{On receiving} \textsc{AgreementNr} %a%\pc{ from \textsc{Executor} }%exr {
		/* Check input%\,%*/
		%\pc{If }%(a <= %\pc{$threshold$}%) return;
		%\pc{If }%(a <= %\pc{$executors$}%[exr]) return;
		%\pc{If }%(a %\pc{is not authentic}%) return;
		
		/* Store input%\,%*/
		%\pc{$executors$}%[exr] := a;
		
		/* Update state%\,%*/
		%\textsc{AgreementNr}% v := %\pc{$executors$}%.highest(F+1);
		%\pc{$threshold$}% := max(%\pc{$threshold$}%, v);
	}

	%\pc{On receiving} \textsc{AgreementNr} %a%\pc{ from \textsc{AgreementMonitor} }%agm {
		/* Check input%\,%*/
		%\pc{If }%(a <= %\pc{$threshold$}%) return;
		%\pc{If }%(a %\pc{is not authentic}%) return;
		
		/* Update state%\,%*/
		%\pc{$threshold$}% := a;
	}
}

class %\textsc{AgreementObserver}% {%\hfill{}%/* Helper Class%\,%*/
	/* State%\,%*/
	%\textsc{NumberOpinions}%<%\textsc{AgreementMonitorID}%, %\textsc{AgreementNr}%> %\pc{$thresholds$}%;
	%\textsc{AgreementNr}% %\pc{$agreed$}%;
	
	/* Initialization%\,%*/
	%\pc{On system start}% {
		%\pc{$thresholds$}% := %\textsc{NumberOpinions}%();
		%\pc{$agreed$}% := 0;
	}
	
	/* Periodic tasks%\,%*/
	%\pc{Periodically}% {
		%\pc{For each \textsc{AgreementMonitor} }%agm: agm.getThreshold();
	}
	
	%\pc{On receiving} \textsc{AgreementNr} %a%\pc{ from \textsc{AgreementMonitor} }%agm {
		/* Check input%\,%*/
		%\pc{If }%(a <= %\pc{$agreed$}%) return;
		%\pc{If }%(a <= %\pc{$thresholds$}%[agm]) return;
		%\pc{If }%(a %\pc{is not authentic}%) return;
		
		/* Store input%\,%*/
		%\pc{$thresholds$}%[agm] := a;
		
		/* Update state%\,%*/
		%\pc{$agreed$}% := %\pc{$thresholds$}%.highest(F+1);
	}
}
\end{lstlisting}

\vspace{-1.5mm}

\subsection{Completion Monitor and Observer}

\vspace{-.5mm}

\begin{lstlisting}[aboveskip=1mm]
class %\textsc{CompletionMonitor}% {
	/* State%\,%*/
	%\textsc{ProgressOpinions}%<%\textsc{ExecutorID}%, %\textsc{CommandProgress}%> %\pc{$executors$}%;
	%\textsc{CommandProgress}% %\pc{$threshold$}%;
	
	/* Initialization%\,%*/
	%\pc{On system start}% {
		%\pc{$executors$}% := %\textsc{ProgressOpinions}%();
		%\pc{$threshold$}% := %\textsc{CommandProgress}%();
	}
	
	/* Services%\,%*/
	%\textsc{CommandProgress}% %\lstbtt{getThreshold}%() {
		return %\pc{$threshold$}%;
	}
	
	/* Periodic tasks%\,%*/
	%\pc{Periodically}% {
		/* Fetch executor thresholds%\,%*/
		%\pc{For each \textsc{Executor} }%exr: exr.getComplete();
		
		/* Fetch monitor thresholds%\,%*/
		%\pc{For each \textsc{CompletionMonitor} }%cpm: cpm.getThreshold();
	}
	
	%\pc{On receiving} \textsc{CommandProgress} %p%\pc{ from \textsc{Executor} }%exr {
		/* Check input%\,%*/
		%\pc{If }%(p %\pc{$\preceq$}% %\pc{$threshold$}%) return;
		%\pc{If }%(p %\pc{$\preceq$}% %\pc{$executors$}%[exr]) return;
		%\pc{If }%(p %\pc{is not authentic}%) return;
		
		/* Store input%\,%*/
		%\pc{$executors$}%[exr] := p;
		
		/* Update state%\,%*/
		%\textsc{CommandProgress}% v := %\pc{$executors$}%.highest(F+1);
		%\pc{For each \textsc{ClientID} }%clt:%\\%%\pc{$threshold$}%[clt] := max(%\pc{$threshold$}%[clt], v[clt]);
	}
	
	%\pc{On receiving} \textsc{CommandProgress} %p%\pc{ from \textsc{CompletionMonitor} }%cpm {
		/* Check input%\,%*/
		%\pc{If }%(p %\pc{$\preceq$}% %\pc{$threshold$}%) return;
		%\pc{If }%(p %\pc{is not authentic}%) return;
		
		/* Update state%\,%*/
		%\pc{For each \textsc{ClientID} }%clt: %\pc{$threshold$}%[clt] := max(%\pc{$threshold$}%[clt], p[clt]);
	}
}

class %\textsc{CompletionObserver}% {%\hfill{}%/* Helper Class%\,%*/
	/* State%\,%*/
	%\textsc{ProgressOpinions}%<%\textsc{CompletionMonitorID}%, %\textsc{CommandProgress}%> %\pc{$thresholds$}%;
	%\textsc{CommandProgress}% %\pc{$complete$}%;
	
	/* Initialization%\,%*/
	%\pc{On system start}% {
		%\pc{$thresholds$}% := %\textsc{ProgressOpinions}%();
		%\pc{$complete$}% := %\textsc{CommandProgress}%();
	}
	
	/* Periodic tasks%\,%*/
	%\pc{Periodically}% {
		%\pc{For each \textsc{CompletionMonitor} }%cpm: cpm.getThreshold();
	}
	
	%\pc{On receiving} \textsc{CommandProgress} %p%\pc{ from \textsc{CompletionMonitor} }%cpm {
		/* Check input%\,%*/
		%\pc{If }%(p %\pc{$\preceq$}% %\pc{$complete$}%) return;
		%\pc{If }%(p %\pc{$\preceq$}% %\pc{$thresholds$}%[cpm]) return;
		%\pc{If }%(p %\pc{is not authentic}%) return;
		
		/* Store input%\,%*/
		%\pc{$thresholds$}%[cpm] := p;
		
		/* Update state%\,%*/
		%\pc{$complete$}% := %\pc{$thresholds$}%.highest(F+1);
	}
}
\end{lstlisting}

\vspace{-1.5mm}

\subsection{View Monitor and Observer}

\vspace{-.5mm}

\begin{lstlisting}[aboveskip=1mm]
class %\textsc{ViewMonitor}% {
	/* State%\,%*/
	%\textsc{NumberOpinions}%<%\textsc{ControllerID}%, %\textsc{View}%> %\pc{$controllers$}%;
	%\textsc{View}% %\pc{$threshold$}%;
	
	/* Initialization%\,%*/
	%\pc{On system start}% {
		%\pc{$controllers$}% := %\textsc{NumberOpinions}%();
		%\pc{$threshold$}% := 0;
	}
	
	/* Services%\,%*/
	%\textsc{View}% %\lstbtt{getThreshold}%() {
		return %\pc{$threshold$}%;
	}
	
	/* Periodic tasks%\,%*/
	%\pc{Periodically}% {
		/* Fetch controller thresholds%\,%*/
		%\pc{For each \textsc{Controller} }%ctr: ctr.getView();
		
		/* Fetch monitor thresholds%\,%*/
		%\pc{For each \textsc{ViewMonitor} }%vwm: vwm.getThreshold();
	}
	
	%\pc{On receiving} \textsc{View} %v%\pc{ from \textsc{Controller} }%ctr {
		/* Check input%\,%*/
		%\pc{If }%(v <= %\pc{$threshold$}%) return;
		%\pc{If }%(v <= %\pc{$controllers$}%[ctr]) return;
		%\pc{If }%(v %\pc{is not authentic}%) return;
		
		/* Store input%\,%*/
		%\pc{$controllers$}%[ctr] := v;
		
		/* Update state%\,%*/
		%\textsc{View}% z := %\pc{$controllers$}%.highest(F+1);
		%\pc{$threshold$}% := max(%\pc{$threshold$}%, z);
	}
	
	%\pc{On receiving} \textsc{View} %v%\pc{ from \textsc{ViewMonitor} }%vwm {
		/* Check input%\,%*/
		%\pc{If }%(v <= %\pc{$threshold$}%) return;
		%\pc{If }%(v %\pc{is not authentic}%) return;
		
		/* Update state%\,%*/
		%\pc{$threshold$}% := v;
	}
}

class %\textsc{ViewObserver}% {%\hfill{}%/* Helper Class%\,%*/
	/* State%\,%*/
	%\textsc{NumberOpinions}%<%\textsc{ViewMonitorID}%, %\textsc{View}%> %\pc{$thresholds$}%;
	%\textsc{View}% %\pc{$view$}%;
	
	/* Initialization%\,%*/
	%\pc{On system start}% {
		%\pc{$thresholds$}% := %\textsc{NumberOpinions}%();
		%\pc{$view$}% := 0;
	}

	/* Periodic tasks%\,%*/
	%\pc{Periodically}% {
		%\pc{For each \textsc{ViewMonitor} }%vwm: vwm.getThreshold();
	}
	
	%\pc{On receiving} \textsc{View} %v%\pc{ from \textsc{ViewMonitor} }%vwm {
		/* Check input%\,%*/
		%\pc{If }%(v <= %\pc{$view$}%) return;
		%\pc{If }%(v <= %\pc{$thresholds$}%[vwm]) return;
		%\pc{If }%(v %\pc{is not authentic}%) return;
		
		/* Store input%\,%*/
		%\pc{$thresholds$}%[vwm] := v;
		
		/* Update state%\,%*/
		%\pc{$view$}% := %\pc{$thresholds$}%.highest(F+1);
	}
}
\end{lstlisting}

\subsection{Adapted Proposer for Shell Committer}
\label{appendix:shellft_base:proposer}

In the following, we present the adapted proposer that is introduced by \shellft if the committer is part of the shell domain, but the proposer itself is not. In this case, the \texttt{VIEW\_CHANGE} mode is replaced to include the stricter checks performed by Mirador's curator and auditor clusters. The remaining functionality of the proposer does not change.

\begin{lstlisting}[aboveskip=1em]
class %\textsc{History}% extends %\textsc{Map}%<%\textsc{CommitterID}%, %\textsc{Legacy}%> {
	/* Operation%\,%*/
	%\textsc{Legacy}% legacy() {
		/* Return if there is no chance of reaching a decision yet%\,%*/
		%\pc{If}% (%\pc{$keys$}%.size() <= 2F+1) return %\pc{$\top$}%;

		/* Determine output%\,%*/
		%\textsc{Legacy}%[] ranking := %\pc{$values$ sorted in descending order of \textsc{Legacy}.$view$}%;
		%\pc{For each \textsc{Number}}% i %\pc{in}% [0, size() - 2F] {
			%\textsc{Number}% acks := 1;
			%\pc{For each \textsc{Number}}% j %\pc{in}% [i + 1, size()] {
				%\pc{If}% (ranking[j].%\pc{$view$}% < ranking[i].%\pc{$view$}%) acks++;
				%\pc{else if}% (ranking[j].%\pc{$command$}% == %\pc{\emptyslot}%) acks++;
				%\pc{else if}% (ranking[j].%\pc{$command$}% == ranking[i].%\pc{$command$}%) acks++;
			}
			%\pc{If}% (acks >= 2F+1) return ranking[i];
		}
		return %\pc{$\top$}%;
	}
}%\vspace{2mm}\hrule\pagebreak\vspace{2mm}%
class %\textsc{Proposer}%$_{Adapted}$ extends %\textsc{AgreementObserver}%, %\textsc{CompletionObserver}%, %\textsc{ViewObserver}% {
	/* Main State%\,%*/
	%\textsc{AgreementWindow}%<%\textsc{Command}%> %\pc{$proposals$}%;
	%\textsc{CommandProgress}% %\pc{$proposed$}%;
	%\textsc{Mode}% %\pc{$mode$}%;%\hfill%/* %\lsttt{NORMAL}%, %\lsttt{VIEW\_CHANGE}%, or %\lsttt{IDLE}\,%*/


	/* Initialization%\,%*/
	%\pc{On system start}% {
		%\pc{$proposals$}% := %\textsc{AgreementWindow}%();
		%\pc{$proposed$}% := %\textsc{CommandProgress}%();
		%\pc{$mode$}% := ite(this == %\pc{$view$}%.%\pc{$proposer$}%, NORMAL, IDLE);
	}
	
	/* Window synchronization%\,%*/
	%\pc{On} \textsc{AgreementNr} \pc{$agreed$ change}% {
		/* Move windows%\,%*/
		%\pc{$proposals$}%.move(%\pc{$agreed$}%);
		shift();%\hfill%/* Mode-specific implementations%\,%*/
	}

	%\pc{On} \textsc{CommandProgress} \pc{$complete$ change}% {
		completed();%\hfill%/* Mode-specific implementations%\,%*/
	}
	
	/* View change%\,%*/
	%\pc{On \textsc{View}}% %\pc{$view$ change}% {
		/* Reset output%\,%*/
		%\pc{$proposals$}%.reset();
		
		/* Switch mode%\,%*/
		%\pc{If }%(this != %\pc{$view$}%.%\pc{$proposer$}%) {
			%\pc{$mode$}% := IDLE;
		} %\pc{else}% {
			%\pc{$proposed$}% := %\pc{$complete$}%;
			%\pc{$mode$}% := VIEW_CHANGE;
		}
	}

	/* Services%\,%*/
	%\textsc{Report}%<%\textsc{Command}%> %\lstbtt{getProposals}%(%\textsc{AgreementRange}% r, %\textsc{View}% v){
		%\pc{If }%((v == %\pc{$view$}%) %\pc{$\wedge$}% (this == v.%\pc{$proposer$}%)) {
			return %\pc{$proposals$}%.report(r, v);
		}
	}%\vspace{1.5mm}\hrule\pc{\begin{center}{\lsttt{NORMAL}}\end{center}}\hrule\vspace*{1.5mm}%
	/* Mode-specific normal-case state%\,%*/
	%\textsc{CommandWindows}%<%\textsc{Command}%> %\pc{$commands$}%;
	
	/* Initialization%\,%*/
	%\pc{On system start}% {
		%\pc{$commands$}% := %\textsc{CommandWindows}%();
	}

	%\pc{On mode start}% {
		/* Move commands%\,%*/
		%\pc{For each \textsc{ClientID}}% clt {
			%\pc{$commands$}%[clt].move(%\pc{$complete$}%[clt]);
		}
		update();
	}
	
	/* Window synchronization%\,%*/
	void shift() {
		update();
	}

	void completed() {
		/* Move commands%\,%*/
		%\pc{For each \textsc{ClientID}}% clt {
			%\pc{$commands$}%[clt].move(%\pc{$complete$}%[clt]);
		}
		update();
	}
	
	/* Periodic command%\,%*/
	%\pc{Periodically}% {
		%\pc{For each \textsc{FrontEnd} }%fre {
			%\textsc{CommandRanges}% rs := %\pc{$commands$}%.empty();
			%\pc{If }%(rs.size() > 0) fre.getCommands(rs);
		}
	}
	
	%\pc{On receiving \textsc{CommandSequence}}%<%\textsc{Command}%>%\,%xss%\pc{ from \textsc{FrontEnd} }%fre {
		/* Store input%\,%*/
		%\pc{For each }%%\textsc{ClientID}% clt {
			/* Check input%\,%*/
			%\pc{If }%(%\pc{$commands$}%[clt].appendable(xss[clt] == false)) return;
			
			/* Check input values%\,%*/
			%\textsc{CommandNr}% from := max(xss[clt].%\pc{$min$}%, %\pc{$commands$}%[clt].%\pc{$pos$}%);
			%\textsc{CommandNr}% to := min(xss[clt].%\pc{$pos$}%, %\pc{$commands$}%[clt].%\pc{$max$}%);
			%\pc{For each }%%\textsc{CommandNr}% x%\pc{ in }%[from, to] {
				%\pc{If }%(xss[clt][x].validate(clt, x) == false) return;
			}

			/* Check input authenticity%\,%*/
			%\pc{If }%(xss %\pc{is not authentic}%) return;

			/* Store input%\,%*/
			%\pc{$commands$}%[clt].append(xss[clt]);
		}
		
		/* Update output%\,%*/
		update();
	}
	
	/* Auxiliary method%\,%*/
	void update() {
		%\pc{For each \textsc{AgreementNr} }%a%\pc{ in }%[%\pc{$proposals$}%.%\pc{$pos$}%, %\pc{$proposals$}%.%\pc{$max$}%] {
			/* Determine output value%\,%*/
			%\textsc{Command}% x := %\pc{nil}%;
			%\pc{For each }%%\textsc{ClientID}% clt%\pc{ in random order}% {
				x := %\pc{$commands$}%[clt][%\pc{$proposed$}%[clt]];
				%\pc{If }%(x != %\pc{nil}%) break;
			}
			%\pc{If }%(x == %\pc{nil}%) break;

			/* Update output value%\,%*/
			%\pc{$proposals$}%[a] := x
			%\pc{$proposed$}%[x.%\pc{$cid$}%] := x.%\pc{$xnr$}% + 1
		}
	}%\vspace{1mm}\hrule\pc{\begin{center}{\lsttt{VIEW\_CHANGE}}\end{center}}\hrule\vspace*{1mm}%
	/* Mode-specific view-change state%\,%*/
	%\textsc{WindowOpinions}%<%\textsc{CommitterID}%, %\textsc{AgreementWindow}%<%\textsc{Legacy}%>> %\pc{$legacies$}%;
	
	/* Initialization%\,%*/
	%\pc{On system start}% {
		%\pc{$legacies$}% := %\textsc{WindowOpinions}%();
	}
	
	%\pc{On mode start or restart}% {
		%\pc{$legacies$}%.sync(%\pc{$proposals$}%);
	}
	
	/* Window synchronization%\,%*/
	void shift() {
		%\pc{$legacies$}%.move(%\pc{$agreed$}%);
	}

	void completed(){
		/* Do nothing%\,%*/
	}

	/* Periodic tasks%\,%*/
	%\pc{Periodically}% {
		%\pc{For each \textsc{Committer} }%cmr {
			%\textsc{AgreementRange}% r := %\pc{$legacies$}%[cmr].empty();
			%\pc{If }%(r.%\pc{$count$}% > 0) cmr.getLegacies(r, %\pc{$view$}%);
		}
	}
	
	%\pc{On receiving \textsc{Report}}%<%\textsc{Legacy}%>%\,%rs%\pc{ from \textsc{Committer} }%cmr {
		/* Check input%\,%*/
		%\pc{If }%(rs.%\pc{$view$}% != %\pc{$view$}%) return;
		%\pc{If }%(%\pc{$legacies$}%[cmr].appendable(rs.%\pc{$values$}%) == false) return;
		%\pc{If }%(rs %\pc{is not authentic}%) return;
		
		/* Store input%\,%*/
		%\pc{$legacies$}%[cmr].append(rs.%\pc{$values$}%);
		
		/* Update output%\,%*/
		%\pc{For each \textsc{AgreementNr} }%a%\pc{ in }%[%\pc{$proposals$}%.%\pc{$pos$}%, %\pc{$proposals$}%.%\pc{$max$}%] {
			/* Determine output value */
			%\textsc{History}% h := %\textsc{History}%();
			%\pc{For each \textsc{CommitterID}}% c {
				%\pc{If}% (%\pc{$legacies$}%[c].%\pc{$pos$}% <= a) continue;
				h[c] := %\pc{$legacies$}%[c][a];
			}
			%\textsc{Legacy}% l := h.legacy();
			%\pc{If}% (l == %\pc{$\top$}%) break;

			/* Switch mode if the output is complete%\,%*/
			%\pc{If }%(l.%\pc{$command$}% == %\pc{\emptyslot}%) {
				mode := NORMAL;
				return;
			}
			
			/* Update output value%\,%*/
			%\textsc{Command}% x := l.%\pc{$command$}%;
			%\pc{$proposals$}%[a] := x;
			%\pc{$proposed$}%[x.%\pc{$cid$}%] := x.%\pc{$xnr$}% + 1;
		}
		
		/* Fill unused input slots%\,%*/
		%\pc{$legacies$}%.fill(%\pc{$proposals$}%.%\pc{$pos$}%, %\pc{\emptyslot}%);
	}%\vspace{2mm}\hrule\pc{\begin{center}{\lsttt{IDLE}}\end{center}}\hrule\vspace*{2mm}%
	/* Window synchronization%\,%*/
	void shift() {
		/* Do nothing%\,%*/
	}

	void completed() {
		/* Do nothing%\,%*/
	}
}
\end{lstlisting}

\end{document}